\documentclass[a4paper,11pt]{article}

\usepackage[totalheight = 23cm, totalwidth = 17cm]{geometry}
\usepackage{amsmath,amssymb,amsfonts,graphicx,bm,physics,subcaption}
\usepackage{rotating,hyperref,slashed,xcolor,colortbl,cite,fancyhdr}

\definecolor{bluc}{cmyk}{1,1,0,0.1}
\definecolor{rossos}{cmyk}{0,1,1,0.55}
\definecolor{verdes}{cmyk}{0.92,0,0.59,0.4}
\hypersetup{colorlinks,bookmarksopen,bookmarksnumbered,
linkcolor=bluc,pdfstartview=FitH,urlcolor=rossos,citecolor=verdes}
\allowdisplaybreaks

\providecommand{\abs}[1]{\lvert#1\rvert} 
\newcommand{\pd}[1]{\partial#1} 
\newcommand{\mpl}{m_{\rm Pl}}

\begin{document}

\begin{titlepage}

\rightline{\footnotesize{APCTP-Pre2022-003}}

\begin{center}

\vskip 3em

{\LARGE \bf 
Effective field theory of waterfall in hybrid inflation
}

\vskip 3em

{\large
Jinn-Ouk Gong$^{a,b}$ 
and
Maria Mylova$^{a}$ 
}

\vskip 0.5cm

{\it
$^{a}$Department of Science Education,  Ewha Womans University, Seoul 03760, Korea
\\
$^{b}$Asia Pacific Center for Theoretical Physics,  Pohang 37673, Korea
}

\end{center}

Email: \href{mailto:jgong@ewha.ac.kr}{jgong@ewha.ac.kr}
and \href{mailto:mmylova@ewha.ac.kr}{mmylova@ewha.ac.kr}

\vskip 1.2cm

\begin{abstract}

We examine the validity of the classical approximation of the waterfall phase transition in hybrid inflation from an effective field theory (EFT) point of view. The EFT is constructed by integrating out the waterfall field fluctuations, up to one-loop order in the perturbative expansion. Assuming slow-roll conditions are obeyed, right after the onset of the waterfall phase, we find the backreaction of the waterfall field fluctuations to the evolution of the system can be dominant. In this case the classical approximation is completely spoiled. We derive the necessary constraint that ensures the validity of the EFT.

\end{abstract}

\end{titlepage}

\newpage

\section{Introduction}
\label{sec:intro}

Currently, inflation \cite{Guth:1980zm,Linde:1981mu,Albrecht:1982wi} -- a phase of accelerated expansion of the universe at early times -- is considered to be the leading candidate to explain otherwise finely tuned initial conditions for the successful hot big bang cosmology. For example, the horizon problem concerns the extremely homogeneous and isotropic temperature of the cosmic microwave background, beyond the causal communication at the moment of the last scattering of the photons. Furthermore, tiny quantum fluctuations during inflation can provide the seeds of observable structure in the universe~\cite{Mukhanov:2005sc,Weinberg:2008zzc,Lyth:2009zz}. The properties of these primordial perturbations have been constrained, during the last decades, by various cosmological observations, with the latest being by the Planck mission~\cite{Planck:2018jri} and are consistent with the predictions of inflation.

To realize a phase of inflation, usually we invoke a hypothetical scalar field, the inflaton, which can provide a negative pressure to cause inflation. Since there is no fundamental scalar field in the standard model of particle physics, a working model of inflation is usually constructed within the theories beyond the standard model~\cite{Lyth:1998xn}. Such theories typically contain a number of scalar fields, e.g. supersymmetric fields and moduli, that can play the role of the inflaton. In this context, hybrid inflation~\cite{Linde:1993cn} has received a particular interest since it is relatively easy to construct this setup within supersymmetric theories~\cite{Linde:1997sj}. The standard, conventional picture of hybrid inflation is as follows: slow-roll inflation is going on until the inflaton reaches a critical value. At that point, another scalar field called the ``waterfall'' field, whose mass is dependent on the inflaton, becomes massless and develops a tachyonic instability, beyond the critical value of the inflaton. The negative mass squared of the waterfall field is increasing very quickly such that, right after the waterfall transition, inflation is terminated almost immediately, as short as less than one single Hubble time.

Under this picture lies the assumption that  the classical approximation is valid during the early stages of waterfall, which translates to the backreaction of the waterfall field fluctuations being initially small. That is, the modes with momentum much smaller than some critical scale, before the waterfall field approaches its true minimum where it starts to oscillate, can well be described by the classical equations of motion. 
Therefore, in this picture, the first stage of the waterfall phase is treated as an effective theory of long-wavelength modes where higher momentum modes have been ``integrated out". These are the quantum modes that remain in the vacuum. Previous methods employed the Langevin equation, where the quantum corrections that were ``integrated out" appear as a correction to the classical dynamics in the form of a noise function \cite{Starobinsky:1986fx,Clesse:2010iz},  or the study of the quantum-to-classical transition, where higher momentum modes where taken to sit in the vacuum state of the system and played no role in the dynamics \cite{Garcia-Bellido:2002fsq}. The latter justifies the use of lattice simulations, which are well suited for the study  of spontaneous symmetry breaking and tachyonic preheating \cite{Felder:2000hj}, as well the use of the classical equations, during the waterfall phase, to study primordial black hole production \cite{Randall:1995dj, Garcia-Bellido:1996mdl, Clesse:2015wea}.

Thus, at the heart of our analytic understanding on the process of waterfall transition in hybrid inflation -- the evolution of the waterfall field can be well described by the classical equation of motion, and so on -- lies the classical approximation. Our aim of this work is to put the classical approximation to the test. As it was pointed above, the classical approximation assumes the backreaction from the waterfall fluctuations is negligible during the first stages of waterfall. This is a reasonable assumption if one focuses mostly on the dynamics of the waterfall field, where one expects the quantum modes that are taken to have been integrated out, can be kept under control till non-linear and non-perturbative processes become important. What has not been entirely clear, so far in the literature, is whether these modes can backreact on the inflaton field during the early stages of the waterfall phase. For example, it is well known that for very flat potentials, in standard chaotic inflation, radiative corrections may spoil the flatness of the potential. Therefore, it is reasonable to wonder whether the backreaction of the waterfall fluctuations could become important.  If this happens, there is danger it will stand the classical approximation invalid.  Indeed, once the inflaton crosses the critical value, we cannot solely rely on the classical solutions any more and a careful investigation is needed.

We study if this is the case by taking a standard field theoretic approach, which has not been so far considered in the literature. It is well known that for an arbitrary number of fields $N$, the waterfall phase in hybrid inflation is well described by a linear sigma model, whose quantum corrections during the symmetry breaking phase have been studied in great detail in the standard text on quantum field theory~\cite{Peskin:1995ev}. Here, we have the added complication that the effective mass squared of the waterfall field changes with time due to the  presence of an interaction term [see \eqref{eq:pp0a}]. For simplicity, on what follows we assume there are only two fields, the inflaton field responsible for inflation and the waterfall field.

The paper is organized as follows: A brief introduction to hybrid inflation can be found in Section~\ref{sec:hybrid}.  Before we embark on our considerations, we give a quick overlook of the various effective field theory (EFT) regimes of hybrid inflation. In Section~\ref{sec:Veff} we integrate out the waterfall fluctuations up to one-loop order and study the dynamics of the system in the presence of quantum corrections. In Section~\ref{sec:SRapprox} we first examine the evolution of the Hubble parameter in the presence of the quantum corrections and find that geometry remains almost de Sitter during the early stages of waterfall. Indeed, in most cases considered in the literature, the rate of the expansion is taken to be constant, once the inflaton  crosses the critical point. We find agreement with this assumption. And then we present the necessary constraint required to maintain control of the EFT, by demanding the backreaction to the inflaton from the waterfall fluctuations be small. This constraint, which has not appeared previously in the literature, ensures the classical description of the system holds so that slow-roll inflation can proceed smoothly and end at the onset of the symmetry breaking phase, as per the standard picture.  Finally, in Section~\ref{sec:num} we provide numerical examples by solving the full system of equations of motion for the inflaton, waterfall field and gravity. We present examples for when the classical approximation is not valid. We also find non-linearities are not important in the regime where our EFT is applicable. We conclude shortly in Section~\ref{sec:conc}. Technical details are relegated to appendices.

\section{Hybrid inflation}
\label{sec:hybrid}

\subsection{Standard picture of hybrid inflation}
\label{hyb}

We start with a brief introduction to hybrid inflation. In what follows and as far as the analytic work is concerned, we do not consider a particular model of hybrid inflation and try to keep things as general and simple as possible. Later in this work, we will investigate numerically specific cases that have been previously considered in the literature.

The action for hybrid inflation is given by
\begin{equation}
\label{eq:pp0}
S =  \int \dd^4{x} a^3 \qty[ \frac{\mpl^2}{2}R - \frac{1}{2} g^{\mu\nu} \pd_\mu \phi \pd_\nu \phi 
- \frac{1}{2} g^{\mu\nu} \pd_\mu \chi \pd_\nu \chi-  V(\phi, \chi) ]
\, .
\end{equation}
Here, the potential is given by
\begin{equation}
V(\phi,\chi) = V_\text{inf}(\phi) 
+ \frac{\lambda}{4} \qty(\frac{M^2}{\lambda} - \chi^2)^2 + \frac{1}{2} g^2 \phi^2 \chi^2
\, ,
\label{eq:pp0a}
\end{equation}
where $\phi (t, {\bm x})$ is the inflaton, $V_\text{inf}$ is the inflaton potential, $\chi (t, {\bm x})$ is the waterfall field, $M$ is the waterfall field mass and $g$ and $\lambda$ are coupling constants. For simplicity, let us assume a quadratic potential for the inflaton:
\begin{equation}
V_\text{inf} = \frac{1}{2} m^2 \phi^2 \, ,
\label{eq:pp1a}
\end{equation}
and work in a Friedmann-Robertson-Walker background space-time:
\begin{equation}
\dd{s}^2=-\dd{t}^2+ a^2(t) \delta_{ij} dx^i dx^j \, .
\label{eq:pp1b}
\end{equation}
From \eqref{eq:pp0}, the equations of motion of the (background) inflaton and waterfall fields are given, respectively, by
\begin{align}
\ddot\phi + 3H\dot\phi + (m^2+g^2\chi^2)\phi & = 0 \, ,
\label{eq:pp1}
\\
\ddot\chi + 3H\dot\chi - \frac{\nabla^2 \chi}{a^2} + M_\text{eff}^2(\phi)\chi + \lambda \chi^3 & = 0 \, ,
\label{eq:pp2}
\end{align}
where the effective mass squared of the waterfall field, which depends on the inflaton field $\phi$, is given by
\begin{equation}
M_\text{eff}^2 \equiv g^2\phi^2 - M^2 \, .
\label{eq:pp3}
\end{equation}
In the regime where $M_\text{eff}^2 >0$, the waterfall field is well anchored at the minimum. During that time, slow-roll inflation takes place, driven by the false vacuum energy $M^4/ (4 \lambda)$. At a critical value of the inflaton field, 
\begin{equation}
\phi_c = \frac{M}{g} \, ,
\label{eq:pp3a}
\end{equation}
we have $M_\text{eff}^2 = 0$. This signals the beginning of the waterfall phase. As the inflaton value decreases, below this point, the waterfall field acquires a negative effective mass squared, $M_\text{eff}^2 < 0$. This is the regime we focus on the rest of this work.

\subsection{Waterfall phase in hybrid inflation}

One can realize the growth of long-wavelength modes by, naively, looking at the dynamical equation for the waterfall field \eqref{eq:pp2}. Assuming that, for simplicity, we are in Minkowski space-time and ignoring the non-linear terms, it can be shown that for $M_\text{eff}^2 <0$, the dispersion of the waterfall field is approximated by~\cite{Felder:2000hj}
\begin{equation}
\expval{\delta \chi^2} 
\approx 
\frac{1}{4 \pi^2} \int^{M_\text{eff}}_0 \dd{k} k 
e^{2t \sqrt{M_\text{eff}^2 - k^2}}
\, ,
\label{eq:pp3b}
\end{equation}
where the integral is taken over the modes with $k < M_\text{eff}$. These are the modes that grow exponentially within $\log (\pi^2/\lambda)/(2m)$, which is understood to be around the region where the double-well potential becomes steep, i.e. where the curvature of the waterfall potential vanishes. Note that this exponential growth of the long-wavelength modes is still observed when the expansion of the universe is taken into account~\cite{Gong:2010zf}. The initial value of $M_\text{eff}$ in \eqref{eq:pp3b} can be estimated by looking in the regime just below the critical value, $\phi < \phi_c$. In the small inflaton regime $M<g$, the effective mass squared of the waterfall field is  $\abs{M_\text{eff}^2} \approx M^2$.

We will later see that the velocity of the inflaton field plays a crucial role in constraining the backreaction of the waterfall field fluctuations during the initial waterfall stage. Indeed, this is not surprising, as it is understood that the transition of $M_\text{eff}^2$ from large and positive to large and negative is controlled by the velocity of the inflaton field at the critical point \cite{Garcia-Bellido:2002fsq}. In standard hybrid inflation \cite{Linde:1993cn}, where the waterfall condition is obeyed, the speed of the inflaton is such that this transition takes place in less than a Hubble time. Usually, the value of the inflaton velocity depends on the model parameters and the energy scale of inflation, therefore, it is treated as an arbitrary parameter.

We find this picture is not entirely correct. The model dependence of the velocity of the inflaton translates to a constraint that excludes a chunk of the parameter space in hybrid inflation, if we are to ensure the validity of the classical approximation and therefore, control of the EFT during the initial waterfall phase. Before we embark on our considerations, it is instructive to first look at the possible EFT regimes in hybrid inflation.

\subsection{EFT regimes in hybrid inflation}

In the following section, we will derive an expression for the effective potential of the waterfall field, for $M_\text{eff}^2 <0$, by integrating out the quantum fluctuations around the top of the double-well potential. In order to understand the regime of validity of this expression it is illustrative to sketch the possible EFT regimes one can obtain in hybrid inflation. These are displayed in Figure~\ref{fig:EFT}. The EFT regimes are as follows:

\begin{figure}[htbp]
\begin{center}
	\includegraphics[width=0.8\textwidth]{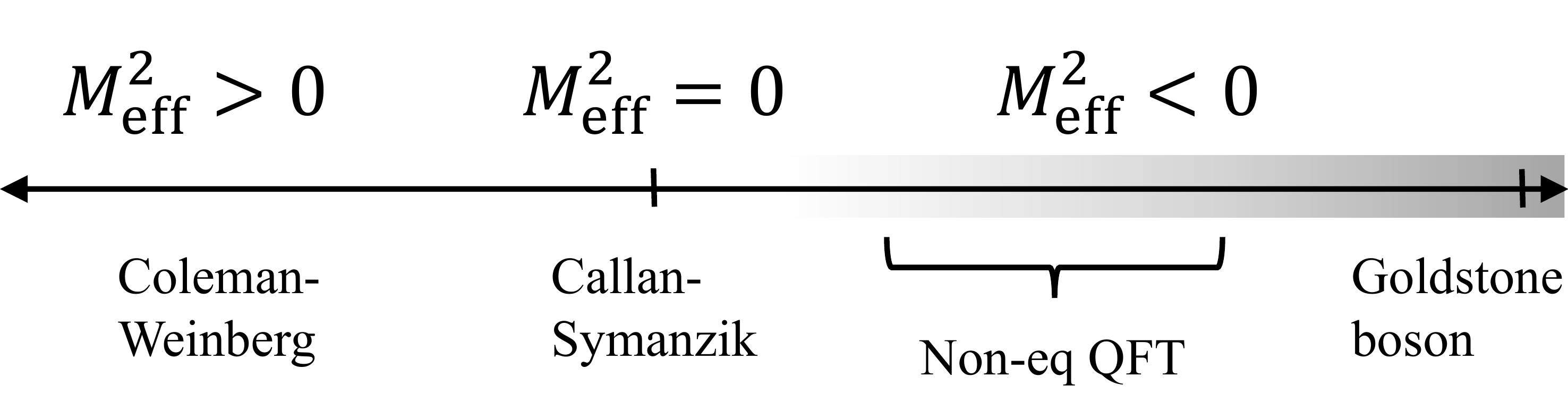}
	\caption{Various EFT regimes of hybrid inflation.}
	\label{fig:EFT}
\end{center}
\setlength{\abovecaptionskip}{0pt plus 1pt minus 1pt}
\end{figure}

\begin{itemize}

\item $\mathbf{M^2_\text{eff} \gg 1}$: This EFT was considered in \cite{Burgess:2003zw}. There, it was shown that one can integrate out the heavy waterfall field, for large positive $M_{eff}^2$. Then one can work instead with the effective potential including the corrections to the dynamics of the inflaton field, that depend logarithmically on the heavy mass $M_\text{eff}^2$ \`a la the Coleman-Weinberg potential~\cite{Coleman:1973jx}. As $M^2_\text{eff}$ becomes much heavier than $H^2$, the waterfall field is more likely to be not excited and we end up with an effective single field inflation where the inflaton is the only dynamical degree of freedom~\cite{Achucarro:2010da,Achucarro:2012sm}.
This EFT is not valid for small values of $M_\text{eff}^2$ and thus can definitely not be used during the waterfall phase.

\item $\mathbf{M^2_\text{eff}=0}$: 
The perturbative expansion around the top of the double-well potential is not valid in the massless limit when $\chi_0 \rightarrow 0$, since a logarithmic term includes a vanishing argument [see \eqref{eq:norm1a}]. The singularity at this point is simply a failure of the computational method and has no physical meaning. Indeed, the correct expression in this limit is given by the Callan-Symanzik resummation of large logarithms \cite{Peskin:1995ev}. In the $N=1$ case, the resulting effective potential $V_\text{eff}$ reads 
\begin{equation}
V_\text{eff}(\chi) = \frac{1}{4} \chi_0^4 \Bigg\{
\bar{\lambda} +  \frac{ 9 \bar{\lambda}^2 }{(4\pi)^2 } \qty[ \log\qty(3 \bar{\lambda} )-\frac{3}{2} ]
\Bigg\} \, ,
\label{eq:w30a} 
\end{equation}
where $\lambda$ is given by
\begin{equation}
\bar{\lambda} = \frac{\lambda}{1-\frac{\lambda}{8 \pi^2} \log(\frac{\chi_0}{\Lambda})} \, .
\label{eq:w31a} 
\end{equation}
As $\chi_0 \rightarrow 0$, $\bar{\lambda} \rightarrow 0$ and so \eqref{eq:w30a} becomes increasingly accurate, ensuring that $V_\text{eff}$ has its minimum at $\chi_0 = 0$. This expression, however, cannot be used to study the dynamics in the waterfall phase, as $M_\text{eff}^2$ quickly becomes massive once $\chi_0$ is displaced from zero. From this, it is understood there is a \textit{discontinuity} in the EFT in the massless limit. This can be understood as the time-dependent mass of the waterfall field becomes small at some region -- when it changes signs and passes through zero -- and at that point it cannot be integrated out. Consequently, the EFT can differ in different regions of the field space.

\item $\mathbf{M^2_\text{eff}<0}$: Here, there are several regimes of validity.

\begin{itemize}

	\item First, we can perturb the waterfall field as $\chi=\chi_0+\delta\chi$ to obtain quantum corrections near the top of the double-well potential, around $\langle \chi \rangle = 0$. These can be integrated out, at quadratic order in $\delta\chi$, to obtain the one-loop corrections to the waterfall potential. At first glance, the effect of these quantum corrections is to shift the value of $\chi_0$ when $V_\text{eff}$ is minimized at zero. In this work, we will use this method to study the early phase of the waterfall transition. There are two interesting limits in this EFT:

	i) In the massless limit $M_\text{eff}^2 \rightarrow 0$, as we briefly discussed above, there is a vanishing log argument for $\chi \rightarrow 0$. This signals the breakdown of the perturbative expansion and large logs have to be resumed using the Callan-Symanzik resummation technique.

	ii) In the limit $\chi_0^2 \rightarrow \abs{ M_\text{eff}^2}/(3 \lambda)$ the logarithmic contributions in the loop corrections blow up, signaling the onset of the non-perturbative waterfall phase. As expected, this is the regime where the effective potential becomes very steep.

	Additionally, the effective potential acquires an imaginary part during the initial stage of the waterfall transition. This will be discussed shortly in more detail.

	\item Once the logs reach non-perturbative values, at about $\chi_0^2 \sim \abs{ M_\text{eff}^2}/(3\lambda)$, it is best to treat the theory numerically by employing non-equilibrium quantum field theory methods \cite{Baacke:2000fw, Baacke:2001zt}. These can be used to study the symmetry breaking phase in detail including non-linear effects.

	\item Finally, the dynamics near the bottom of the potential, for an $SO(N)$ theory, can be expressed in  terms of the Goldstone bosons. This description of the system is only valid when expanding about the true vacuum expectation value of the field.

\end{itemize}

\end{itemize}

\section{Effective potential}
\label{sec:Veff}

We are interested in the one-loop corrections to the waterfall potential at the early stage of the waterfall transition, i.e. once $\phi < \phi_c$. The Lagrangian of the waterfall field is described by 
\begin{equation}
\mathcal{L} =  -\frac{1}{2} g^{\mu\nu} \pd_\mu \chi \pd_\nu \chi 
+ \frac{1}{2} M^2_\text{eff}(\phi) \chi^2 +  \frac{\lambda}{4} \chi^4 + \text{counterterms}
\label{eq:w4}
\end{equation}
for $M^2_\text{eff}(\phi) <0$, where the counterterms are to be determined, order by order, in the loop expansion parameter. This action has a discrete\footnote{One could instead consider an $SO(N)$ scalar model, also known as the linear sigma model. Then, there would be two types of loop diagrams to consider. In any case, our results should be the same as taking $N=1$ in the end. Also, note that the Goldstone's theorem does not apply here, since we do not consider a continuous symmetry.} $Z_2$ symmetry $\chi \rightarrow - \chi$. The potential has a false minimum at $\chi_0=0$ and a vacuum expectation value at $\chi_0 = \pm \abs{M_\text{eff}}/ \sqrt{\lambda}$.

We expand the waterfall field around some classical value as $\chi = \chi_0 + \delta\chi$, where $\langle \chi \rangle = \chi_0$ is some classical configuration of the waterfall field and $\delta\chi$ describes the quantum corrections around $\chi_0$. The details of this calculation can be found in Appendix~\ref{app:int-out}. From this, we find that the effective potential for the waterfall field, up to one-loop corrections, is given by
\begin{equation}
V_\text{eff}(\chi) 
= 
- \frac{1}{2}\abs{ M^2_\text{eff}} \chi_0^2 +  \frac{\lambda}{4} \chi_0^4 
+ \frac{1}{4} \frac{\qty(-\abs{M_\text{eff}^2} + 3 \lambda \chi_0^2)^2}{(4\pi)^2 } 
\qty[ \log \qty( \frac{-\abs{M_\text{eff}^2 }+ 3 \lambda \chi_0^2}{\Lambda^2} ) - \frac{3}{2} ]
\, ,
\label{eq:norm1a} 
\end{equation}
where we have made it explicit that the negative effective mass squared is $M_\text{eff}^2 = - \abs{M_\text{eff}^2}$ and $\Lambda$ is an arbitrary renormalization scale. From \eqref{eq:norm1a}, we see that for the EFT to be valid, it requires 
\begin{equation}
\abs{\chi_0} <  \abs{\frac{ M_\text{eff}}{\sqrt{3 \lambda}}} \, .
\label{eq:i2}
\end{equation}
Once $\chi_0$ reaches this value, the perturbative expansion is no longer valid and the EFT loses its predictability.  Beyond this regime lies the minimum of the effective potential at $\chi_0 = \pm \abs{M_\text{eff}} / \sqrt{\lambda}$.

\begin{figure}
	\begin{subfigure}[b]{0.33\textwidth}
		\centering
		\includegraphics[width=\linewidth]{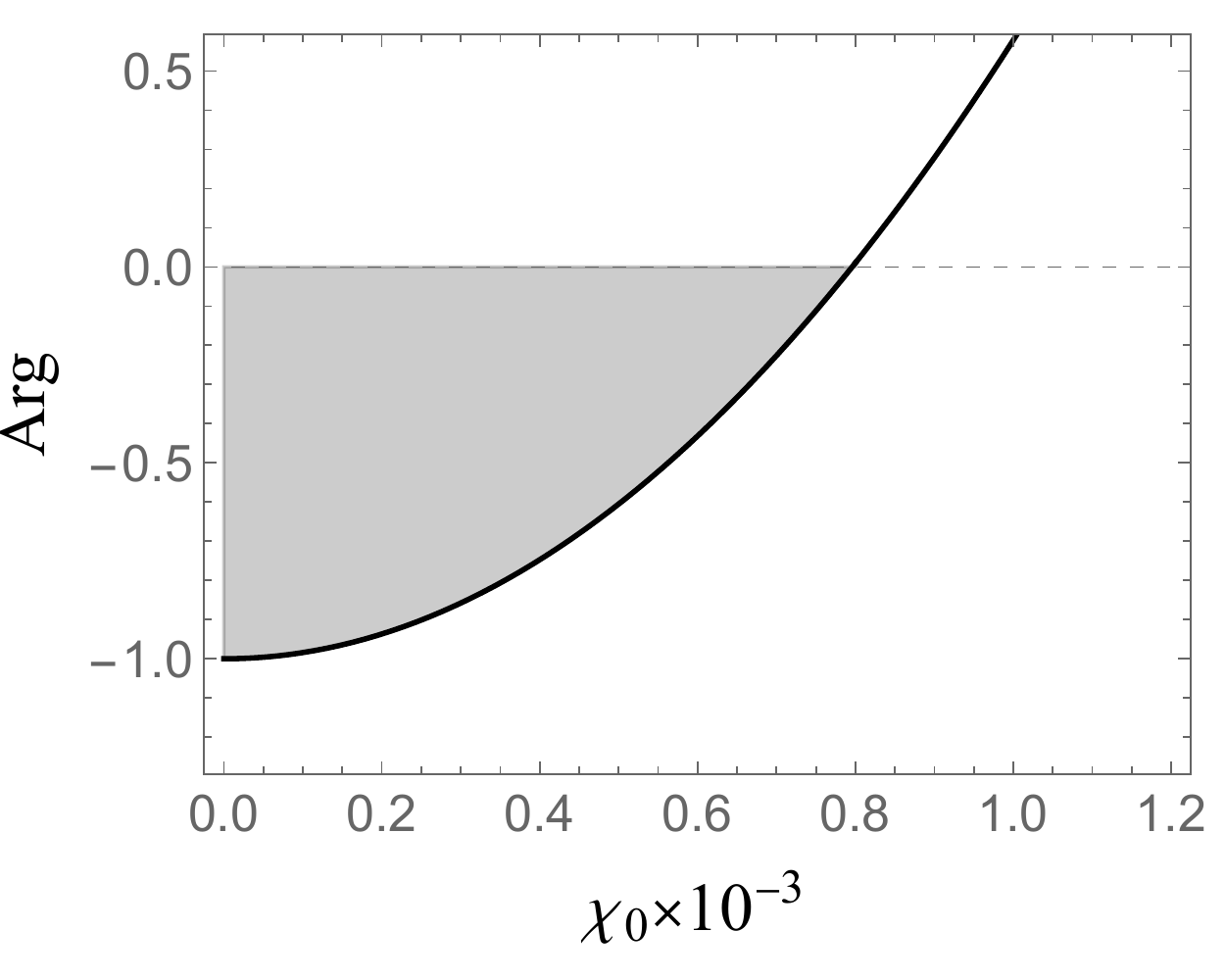}
		\caption{The argument of the log in \eqref{eq:norm1a}}
		\label{fig:loga}
	\end{subfigure}\hfill
	\begin{subfigure}[b]{0.33\textwidth}
		\centering
		\includegraphics[width=\linewidth]{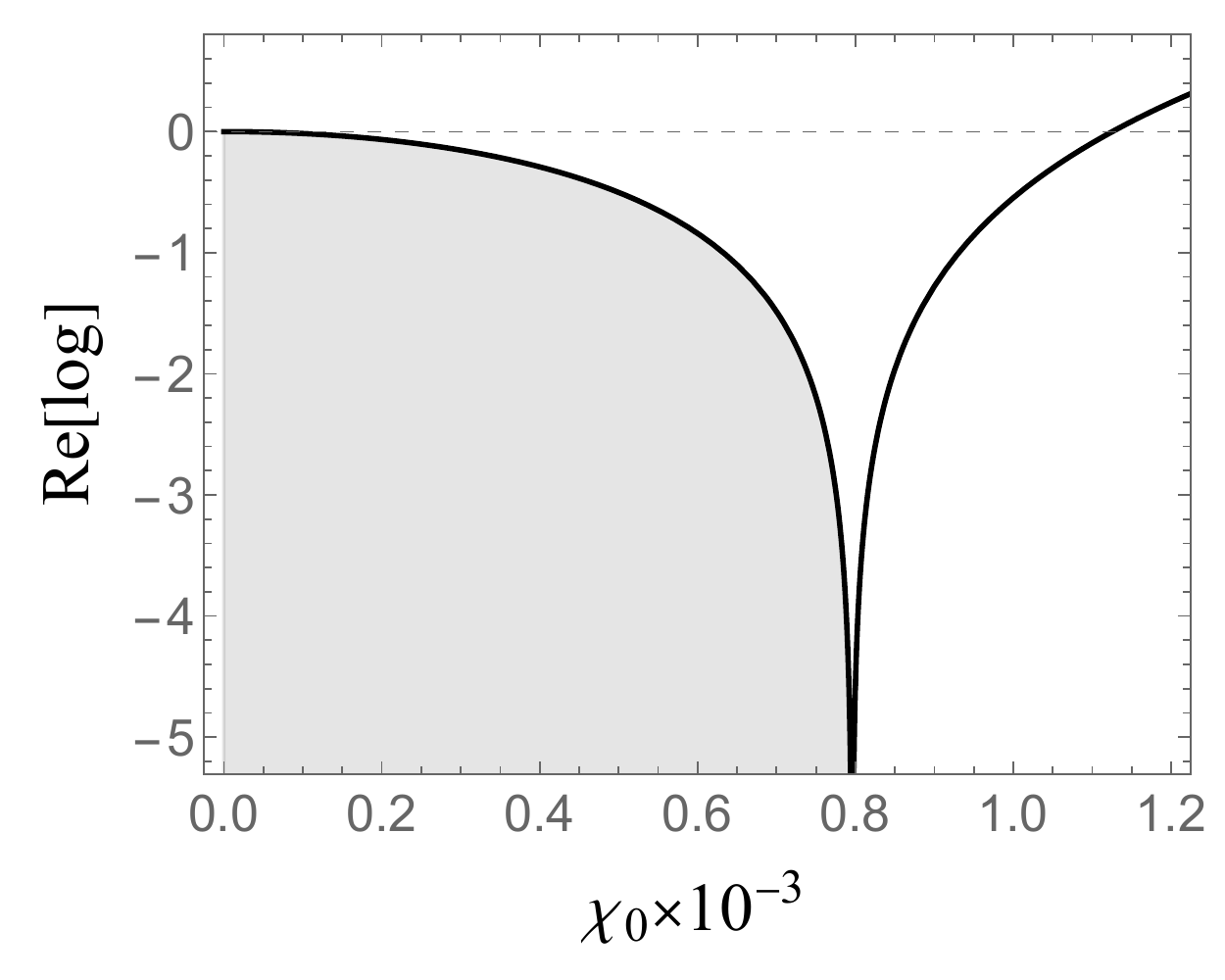}
		\caption{Real component of the log}
		\label{fig:logb}
	\end{subfigure}
	\begin{subfigure}[b]{0.33\textwidth}
		\centering
		\includegraphics[width=\linewidth]{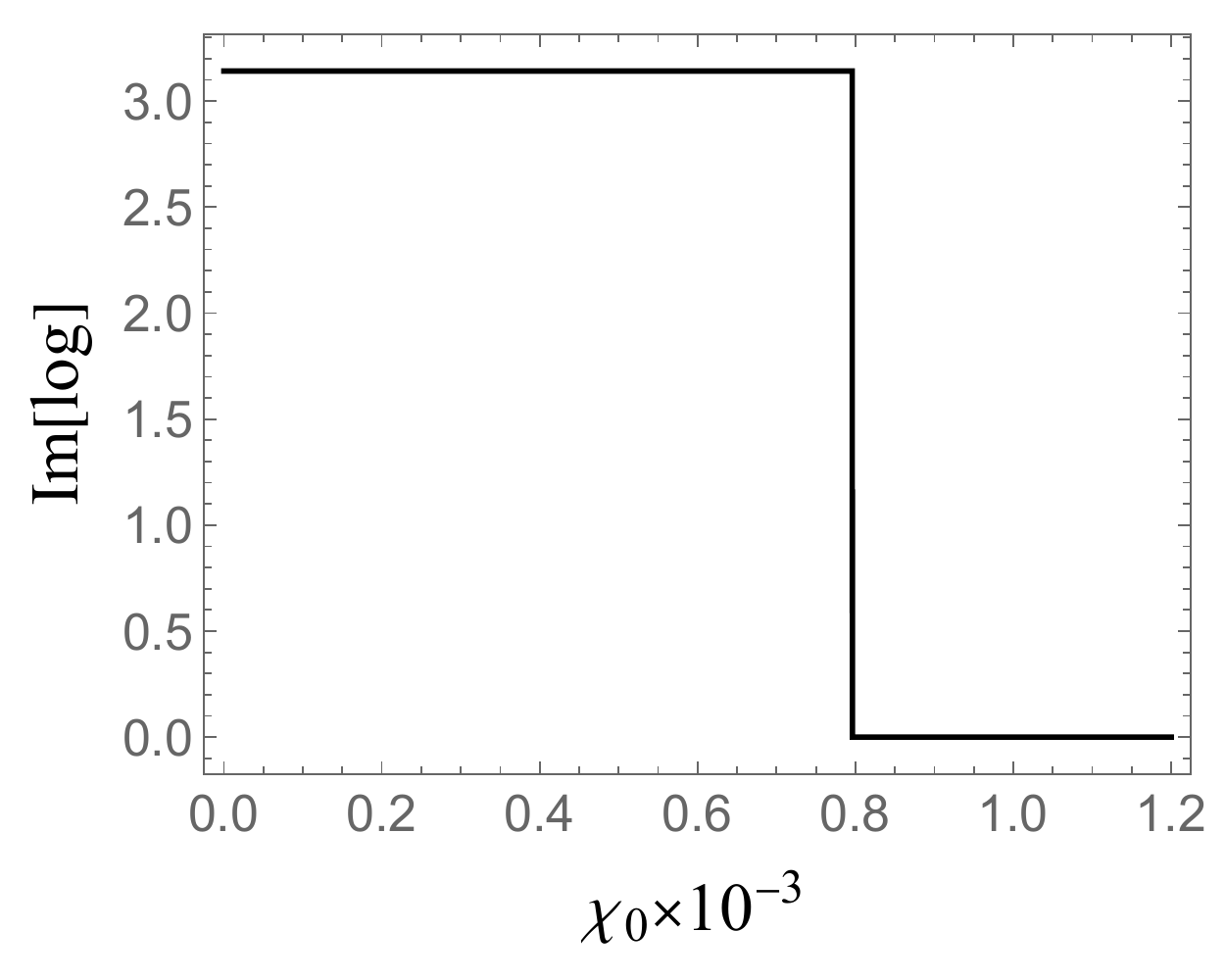}
		\caption{Imaginary component of the log}
		\label{fig:logc}
	\end{subfigure}
	\caption{Plotting the log contribution in \eqref{eq:norm1a}, starting at $\chi_0\approx0$. The perturbative validity of our expansion is indicated by the shaded region. Perturbation theory breaks down for $\abs{\chi_0} \rightarrow   \abs{M_\text{eff}}/ \sqrt{3 \lambda}$. Here, we have chosen $\Lambda^2 \sim \abs{M_{eff}^2}$ such  that the real component of the argument of the logarithm at $\chi_0=0$ is zero.}
	\label{fig:log}
\end{figure}

\subsection{Imaginary corrections to the effective potential}

Next, we focus on the logarithmic contribution. Looking at Figure~\ref{fig:log}, we see that during the initial phase of the waterfall, indicated by imposing the constraint \eqref{eq:i2}, the argument of the logarithm is negative and the effective potential contains imaginary contributions. This is the regime where we will focus in the rest of this text.  Using $\log x=i\pi + \log \abs{x}$ for $x<0$, \eqref{eq:norm1a} can be written as 
\begin{equation}
V_\text{eff}(\chi) = - \frac{1}{2}   \abs{ M_\text{eff}}^2 \chi_0^2 +  \frac{\lambda}{4} \chi_0^4 
+\frac{1}{4} \frac{\qty( - \abs{ M_\text{eff}}^2 + 3 \lambda \chi_0^2)^2}{(4\pi)^2 }
\qty[ i\pi + \log \qty(\frac{  \abs{ M_\text{eff}}^2 - 3 \lambda \chi_0^2 }{\Lambda^2})-\frac{3}{2}]
\, .
\label{eq:i3} 
\end{equation}
From this, we see the imaginary corrections to $V_\text{eff}$ are proportional to
\begin{equation}
\frac{i \pi}{4} \frac{\qty( - \abs{ M_\text{eff}} + 3 \lambda \chi_0^2)^2}{(4\pi)^2 }
\, ,
\label{eq:i6a} 
\end{equation}
and therefore small as long as \eqref{eq:i2} is obeyed. This is required for the perturbation theory to be valid.

At this point one may object that the theory contains particles with imaginary mass. Indeed, we are doing perturbation theory as if it did, but this is acceptable as we only work to finite order in the loop expansion.  We treat this strictly as an EFT and it must be understood that unitarity will be restored if we were to calculate the imaginary part to all orders. Indeed, the loss of unitarity can be seen through the manifestation of an imaginary part to the effective potential and this can be related to the vacuum decay rate. All of the above considerations are as per the ancient lore in complex effective potentials from particle physics, and the interested reader can look at \cite{Weinberg:1987vp} and at the excellent introduction to EFT \cite{Burgess:2020tbq} for another example with positronium decay.

What is different from the standard lore is that the waterfall field is coupled not only to the inflaton but also to gravity. This means that through the system of coupled equations of motion, both the inflaton field and the Hubble parameter will obtain an imaginary part. This is not to be confused with that the energy density of the universe, during that time, is not real. But this is simply to be understood as a loss of probability which can be related to the vacuum decay rate. Indeed, this is to be expected as the waterfall field undergoes through a process that leads to spontaneous symmetry breaking. To put it more plainly, the EFT is simply a perturbative expansion in powers of the coupling constant $\lambda$. Therefore, it fails to account for the non-perturbative process of vacuum decay. This failure manifests itself as an imaginary part in the effective potential.

In this work, we study what happens in the first moments of the waterfall transition by always staying within the perturbative validity of the EFT. Therefore, we require the backreaction of the waterfall fluctuations, both real and imaginary contributions, to the inflaton field and gravity be small, for as long as \eqref{eq:i2} is obeyed.

Note that the minimum of the effective potential is not affected by the imaginary part, as it vanishes once $\abs{\chi_0} >  \abs{ M_\text{eff}}/\sqrt{3 \lambda}$ (see Figure~\ref{fig:log}). Nevertheless, this stage of the waterfall process is beyond the validity of the EFT due to the constraint \eqref{eq:i2}.

\subsection{Corrections to the equations of motion}
\label{EOM}

Using the effective potential \eqref{eq:i3}, it becomes manifest how the real and imaginary components in \eqref{eq:i3} affect the system of equations to solve. We have the following equations of motion for the inflaton field $\phi$, the background waterfall field $\chi_0$ and the Friedmann equation:
\begin{align}
& 
\ddot\phi + 3 H \dot\phi + \qty( m^2+ g^2 \chi_0^2 ) \phi 
+ \frac{g^2 \phi }{32 \pi^2 } \qty( -\abs{M_\text{eff}^2} + 3 \lambda \chi_0^2 ) 
\nonumber\\
& 
+ \frac{g^2 \phi}{16 \pi^2} \qty( -\abs{M_\text{eff}^2} + 3 \lambda \chi_0^2) 
\qty[ \log \qty(\frac{\abs{M_\text{eff}^2}- 3 \lambda \chi_0^2}{\Lambda^2})+ i \pi-\frac{3}{2}] = 0
\, ,
\label{eq:i11} 
\\
& 
\ddot\chi_0 + 3 H \dot\chi_0 - \abs{M_\text{eff}^2} \chi_0 + \lambda \chi_0^3 
+ \frac{3 \lambda  \chi_0 }{32 \pi^2 } \qty(-\abs{M_\text{eff}^2} + 3 \lambda \chi_0^2) 
\nonumber\\
& 
+ \frac{3 \lambda \chi_0}{16 \pi^2} \qty(-\abs{M_\text{eff}^2} + 3 \lambda \chi_0^2) 
\qty[ \log \qty(\frac{\abs{M_\text{eff}^2}- 3 \lambda \chi_0^2}{\Lambda^2})+ i \pi-\frac{3}{2}] =0
\, ,
\label{eq:i12} 
\\
H^2 =
& 
\frac{1}{3\mpl^2} \Bigg\{ \frac{\dot\phi^2}{2}+ \frac{\dot\chi_0^2}{2} 
+ \frac{M^4}{4 \lambda} + \frac{m^2 \phi^2}{2} 
- \frac{1}{2}   \abs{ M_\text{eff}}^2 \chi_0^2 +  \frac{\lambda}{4} \chi_0^4
\nonumber\\
&  
\hspace{3em}
+ \frac{1}{4} \frac{\qty( - \abs{ M_\text{eff}}^2 + 3 \lambda \chi_0^2)^2}{(4\pi)^2 } 
\qty[ i\pi + \log \qty(\frac{  \abs{ M_\text{eff}}^2 - 3 \lambda \chi_0^2 }{\Lambda^2})-\frac{3}{2} ] \Bigg\}
\, .
\label{eq:i14} 
\end{align}
We can solve this system of equations numerically to find the behaviour of the solutions for a large range of initial conditions and parameters.  Before doing so, we will examine analytically the validity of the EFT in the slow-roll approximation. This is the subject of the next section.

Note the above expressions seem to depend on the choice of the renormalisation scale $\Lambda$. The value of this scale is arbitrary and therefore, for our computational purposes, we choose it to have a value such that the argument of the logarithm is initially unity. Indeed, it is realistic to use very small values for the initial value of $\chi_0 < m$, which means that $\Lambda$ has almost the same value as $M_\text{eff}^2$ for $\chi_0 \rightarrow 0$ just below $\phi_c$, ensuring small logs are part of our initial conditions. This is a sensible choice, as $\Lambda$ is expected to be the same order as the effective mass scale of the waterfall field. Additionally,  ensuring small logs as part of our initial conditions provides agreement, during the first part of the evolution of the fields, with the Callan-Symanzik expressions discussed earlier.

\section{Slow-roll approximation during waterfall}
\label{sec:SRapprox}

\subsection{The role of gravity during waterfall}
\label{grav}

Before we embark on our computations it is necessary to ensure the space-time geometry stays unaffected in the presence of the waterfall quantum corrections. Therefore, assuming the slow-roll approximation, we examine the backreaction of the waterfall field fluctuations to gravity. Our starting point is the Friedmann equation. In the following, it is convenient to get rid of the logs in the effective potential by Taylor expanding the logs as $\log x = x-1$. This enables us to use much simpler expressions. For details, see Appendix~\ref{app:Taylor-log}. The Hubble parameter in the slow-roll approximation is given by
\begin{equation}
H = \frac{1}{\sqrt{3}\mpl} \sqrt{ \frac{M^4}{4 \lambda}+\frac{1}{2} m^2 \phi^2 +V_\text{eff}(\phi,\chi_0)} \, ,
\label{eq:gsr1}
\end{equation}
where $V_\text{eff}(\phi,\chi_0)$ is given by \eqref{eq:i7}. The expression under the square root is simply a complex function of the form $z= x+ i y$. Therefore, we can separate the real and imaginary parts by employing De Moivre's theorem, giving 
\begin{equation}
H 
= 
\frac{1}{\sqrt{3}\mpl} \abs{z}^{1/2} \Bigg\{ \cos \qty[\frac{1}{2} \arg(z)] + i \sin \qty[\frac{1}{2} \arg(z)] \Bigg\} \, ,
\label{eq:gsr4}
\end{equation}
where the modulus of $z$ is given by $\abs{z} = \sqrt{x^2+y^2}$ and the argument of $z$ is given by the two-argument inverse tangent function $\arctan{(y/x)}$, whose precise value depends on the signs of $x$ and $y$. Next, we use that during slow-roll inflation the inflaton is almost constant therefore, the dynamical behaviour in \eqref{eq:gsr4} is predominantly due to the waterfall field evolving towards its true minimum.  To understand if the imaginary part becomes important during this process, it suffices to study the cosine and sine functions at different stages in the evolution of $\chi_0$.

It is best to express the argument of $z$ in a more useful form that will enable us to study which contributions in the effective potential are relevant during this process. We can write 
\begin{equation}
\arg{z} = \arg\qty(a_0+a_1 \chi_0^2 + a_2 \chi_0^4  + a_3 \chi_0^6),
\label{eq:gsr5}
\end{equation}
where the complex constant coefficients $a_0$, $a_1$, $a_2$ and $a_3$ are given below: 
\begin{align}
a_0 
& =
\frac{M^4}{4 \lambda }+\frac{m^2\phi ^2}{2}-\frac{5 M^4}{128 \pi ^2}+\frac{M^6}{64 \pi ^2 \Lambda ^2}
+\frac{5 g^2 M^2 \phi ^2}{64 \pi ^2}-\frac{3 g^2 M^4 \phi ^2}{64 \pi ^2 \Lambda^2}
-\frac{5 g^4 \phi ^4}{128 \pi ^2}+\frac{3 g^4 M^2 \phi ^4}{64 \pi ^2 \Lambda^2}
-\frac{g^6 \phi ^6}{64 \pi ^2 \Lambda ^2}
\nonumber\\
& \quad
+ i \left(\frac{M^4}{64 \pi }-\frac{g^2 M^2 \phi ^2}{32 \pi }+\frac{g^4 \phi ^4}{64 \pi }\right)
\, ,
\label{eq:gsr6}
\\
a_1 
& =
-\frac{M^2}{2}+\frac{15 M^2 \lambda }{64 \pi ^2}-\frac{9 M^4 \lambda }{64 \pi ^2 \Lambda^2}
+\frac{g^2 \phi ^2}{2}-\frac{15 g^2 \lambda  \phi ^2}{64 \pi ^2}
+\frac{9 g^2 M^2\lambda  \phi ^2}{32 \pi ^2 \Lambda ^2}-\frac{9 g^4 \lambda  \phi ^4}{64 \pi ^2 \Lambda^2}
+ i \left(-\frac{3 M^2 \lambda }{32 \pi }+\frac{3 g^2 \lambda  \phi ^2}{32 \pi }\right)
\, ,
\label{eq:gsr7}
\\
a_2 
& =
\frac{\lambda }{4}-\frac{45 \lambda ^2}{128 \pi ^2}+\frac{27M^2 \lambda ^2}{64 \pi ^2 \Lambda ^2}
-\frac{27 g^2 \lambda ^2 \phi ^2}{64 \pi ^2 \Lambda^2} 
+i \frac{9 \lambda ^2}{64 \pi }
\, ,
\\
a_3 
& = 
-\frac{27 \lambda ^3}{64 \pi ^2 \Lambda ^2}
\, .
\label{eq:gsr8}
\end{align}
As far as the initial conditions for the Hubble parameter are concerned, in the slow-roll approximation and in the small inflaton regime with $\phi<\phi_c$, we find the coefficient $a_0$ dominates for the initial evolution of  waterfall phase, as the other coefficients are suppressed by the small initial value of the waterfall field. The real part of $a_0$ is always much larger than the imaginary part due to that the leading-order contribution $M^4/(4 \lambda)$ dominates over the quantum corrections, as expected. 
As this term is, however, practically very small compared to $\mpl^4$, when working in units of $\mpl=1$, we have $\arg z \ll 1 $ and the cosine function evaluates to almost unity, $\cos [\arg (z)/2] \approx 1$, while the sine function is almost zero during that time, $\sin [\arg (z)/2] \ll 1$.

Therefore, ensuring $a_0$ dominates as part of the initial conditions indicates the backreaction of the waterfall field fluctuations to gravity is initially small. Once the waterfall field starts to evolve, there are two possibilities:

\begin{itemize}

\item Backreaction to gravity: The waterfall field will, eventually,  evolve towards its minimum and the $\chi_0^2$, $\chi_0^4$ and $\chi_0^6$ terms will become dynamical. If they become dominant, these additional contributions may lead to $\abs{\arg z} > 1$. Consequently the cosine function may approach $\cos [\arg (z)/2] \sim 1/2$, while the sine function may approach unity. In this case the imaginary part of the Hubble parameter can become almost equal to the real part. So far, we have not found any evidence of backreaction to gravity while staying within the validity of the EFT. This translates to the geometry being almost de Sitter, as expected.

\item In general, the waterfall field is expected to reach its minimum\footnote{This is, of course, oversimplified. In truth slow-roll inflation is expected to end once the waterfall reaches the steep part of the double-well potential. In ordinary hybrid inflation this picture depends on the relation between the coupling constants $\lambda$ and $g$.} just as slow-roll inflation ends. Due to that the slow-roll approximation is no longer valid, the coefficients $a_0$, $a_1$, $a_2$ and $a_3$ will become dynamical, as they are functions of the inflaton field. We find, once this happens, the imaginary part of the Hubble parameter can grow rapidly and we will lose control of the EFT.  This is fine, as our perturbative expansion is around the top of the double-well potential and therefore we do not expect to have reliable predictions beyond the point where the effective potential becomes steep. The non-perturbative process of vacuum decay, including the effects of gravity, was previously examined in \cite{Coleman:1980aw}.

\end{itemize}
To summarize this section, we have derived a fully analytic expression for the Hubble parameter \eqref{eq:gsr4} which can be used to study the backreaction of the waterfall field fluctuations to gravity during the early stages of the waterfall transition. So far, we have found no evidence of large backreaction to gravity, while staying within the validity of the EFT. Therefore, the space-time geometry stays unaffected during the initial waterfall stage.

\subsection{Controlling the backreaction of the waterfall field fluctuations}
\label{backr}

Assuming the slow-roll approximation is obeyed during the early stages of the waterfall phase, we can derive an analytic expression for the slow-roll parameter by using 
\begin{equation}
\epsilon 
= 
\frac{\dot\phi^2+\dot\chi_0^2}{2 \mpl^2 H^2} 
= 
\frac{V_{,\phi}^2+V_{,\chi_0}^2}{18 \mpl^2 H^4}
\, ,
\label{eq:sl1} 
\end{equation}
where $V_{,\phi}$ and $V_{,\chi_0}$ denote, respectively, the differentiation of the effective potential with respect to the inflaton and to the waterfall field.

To maintain slow-roll inflation during the early stages of waterfall, it requires $\Re \epsilon \ll 1$ and $\Im \epsilon \ll \Re \epsilon$. The last condition is simply the requirement that the backreaction of the waterfall field fluctuations is small. As we discussed in the previous section, the space-time geometry is not likely to be affected during this process and therefore it is safe to take $H$ being constant. Looking  at expression \eqref{eq:sl1}, we see that the only danger for breaking the above requirements on the slow-roll parameter $\epsilon$ is whether the backreaction of the waterfall field fluctuations dominates the inflaton field dynamics. Let us then look at the conditions that ensure the velocity of the inflaton field lies within the validity of the EFT.

Here, we employ initial conditions such that we can discard the logs, as we choose the scale $\Lambda$ so that the logs are vanishingly small as a part of our initial conditions. This can be simplified further by imposing very small values for the initial value of the waterfall field\footnote{For realistic initial conditions one needs to demand initially $\chi_0 < m$, subject to the mass hierarchy $m<H<M$.}. Indeed, we find that the initial value of $\chi_0$ does not play an important role, the only requirement being that it is sufficiently small. With these considerations, the expression for the inflaton velocity simplifies significantly. We have
\begin{equation}
\dot\phi 
= 
\frac{1}{3 H}\qty[m^2 \phi + \frac{g^2 \phi}{16\pi^2} \abs{ M^2_\text{eff}}  \qty( 1 - i\pi ) ] \, .
\label{eq:sl4}
\end{equation}
Looking at the contributions inside the square brackets, the first term is coming from the quadratic inflaton potential, while the rest terms give the quantum corrections to the inflaton velocity. We see that the waterfall field fluctuations can backreact strongly on the inflaton if the mass of the inflaton is too small and the coupling constant $g$ is too large. Therefore, we need to constrain the ratio $m/g$ according to the following:
\begin{equation}
\frac{m}{g} > \frac{\abs{M_\text{eff}}}{4 \sqrt{\pi}} \approx \frac{M}{4 \sqrt{\pi}} \, . 
\label{eq:sl7} 
\end{equation}
Here, we have used that $M_\text{eff} \approx M$ during the waterfall stage, which is reasonable in the small inflaton regime. This is the necessary condition required for the classical approximation to be valid. That is, for the waterfall field fluctuations that were integrated out to not become important and spoil the validity of the EFT. This was not taken into consideration in previous works in the literature.

This constraint means that the mass of the inflaton field cannot be too small. This indicates that the backreaction can be serious when the inflaton potential is taken to be very flat and it is reminiscent to the $\eta$ problem in generic inflation models in supergravity~\cite{Copeland:1994vg}. Additionally, we find the coupling constant $g$ between the inflaton and the waterfall field cannot be too large. Indeed, a suitable choice of $m$ and $g$ can ensure slow-roll inflation proceeds during the waterfall phase while the classical approximation is under control.

Later, when we examine the evolution of the system numerically, we find that when this condition is not obeyed, the quantum fluctuations become large and the EFT of the coupled inflaton-waterfall system breaks down already at the onset of the  waterfall phase transition. This failure of the EFT manifests itself through large real and imaginary parts to the inflaton velocity and slow-roll parameter, coming from the quantum corrections, that cannot be neglected.

Finally, as we mentioned earlier, the velocity of the inflaton controls how quickly the effective mass squared of the waterfall field changes from large and positive to large and negative therefore, constraining the backreaction of the waterfall field fluctuations to the inflaton dynamics. This naturally translates to a constraint on the velocity of the inflaton.

\section{Numerical results}
\label{sec:num}

For the numerical calculations, we choose the initial value of the inflaton field to be just below the critical value. This is because our EFT is strictly valid during the waterfall phase and not before.  For some of the parameters, we use initial values with several significant figures which will not be displayed here.

We would like to add that one may choose to study the real and imaginary part of the system of equations \eqref{eq:i14} separately. This is of course just a matter of preference, as the numerical methods we use are perfectly capable of separating real and imaginary contributions in an appropriate way. The results are the same, whichever way one performs these calculations.  Here, for convenience, we choose the latter.

\subsection{Hybrid inflation with two stages of slow-roll phase}

This case was originally studied  in \cite{Clesse:2010iz,Kodama:2011vs}. We numerically solve the system of equations \eqref{eq:i11}, \eqref{eq:i12} and \eqref{eq:i14} using the following initial values (in the unit $\mpl=1$) for the parameters, taken from \cite{Clesse:2010iz}:
\begin{equation}\begin{split} 
& \lambda = 5\times 10^{-14}, \quad  g = 2\times 10^{-7}, \quad M=3 \times 10^{-8}, \quad m = 8\times 10^{-12},
\\
& \chi_0(t=0) = 10^{-12}, \quad \phi(t=0)= 1.4 \times 10^{-1}, \quad \Lambda \sim 1.1 \times 10^{-8}.
\label{eq:n1a} 
\end{split}\end{equation}
As noted earlier, our system of equations is applicable in the regime $\phi < \phi_c$. Using the values above we have $M/g = 0.15$, so we start the computation just below that value. This, consequently, fixes the value of $\Lambda$, so to ensure small logs as part of the initial conditions such  that $\log[\phi(t=0), \chi_0(t=0)]=\log1=0$. This is a reasonable condition which ensures $\Lambda$ is of the same order as $M_\text{eff}$.  Once the fields start evolving, of course, the logs become non-vanishing and contribute to the computations.

\begin{figure}
	\begin{subfigure}[b]{0.33\textwidth}
		\centering
		\includegraphics[width=\linewidth]{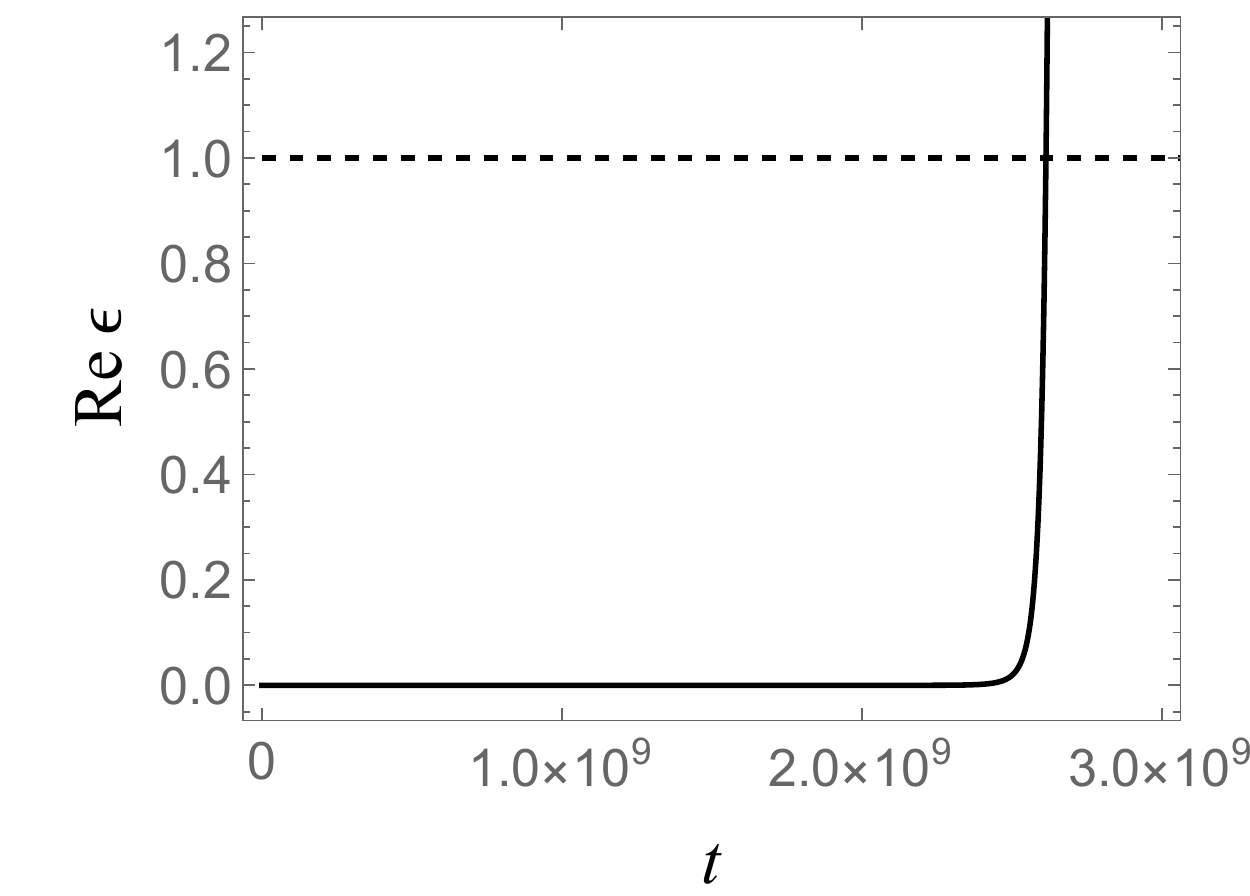}
		\caption{$\Re \epsilon$}
		\label{fig:ca}
	\end{subfigure}\hfill
	\begin{subfigure}[b]{0.33\textwidth}
		\centering
		\includegraphics[width=\linewidth]{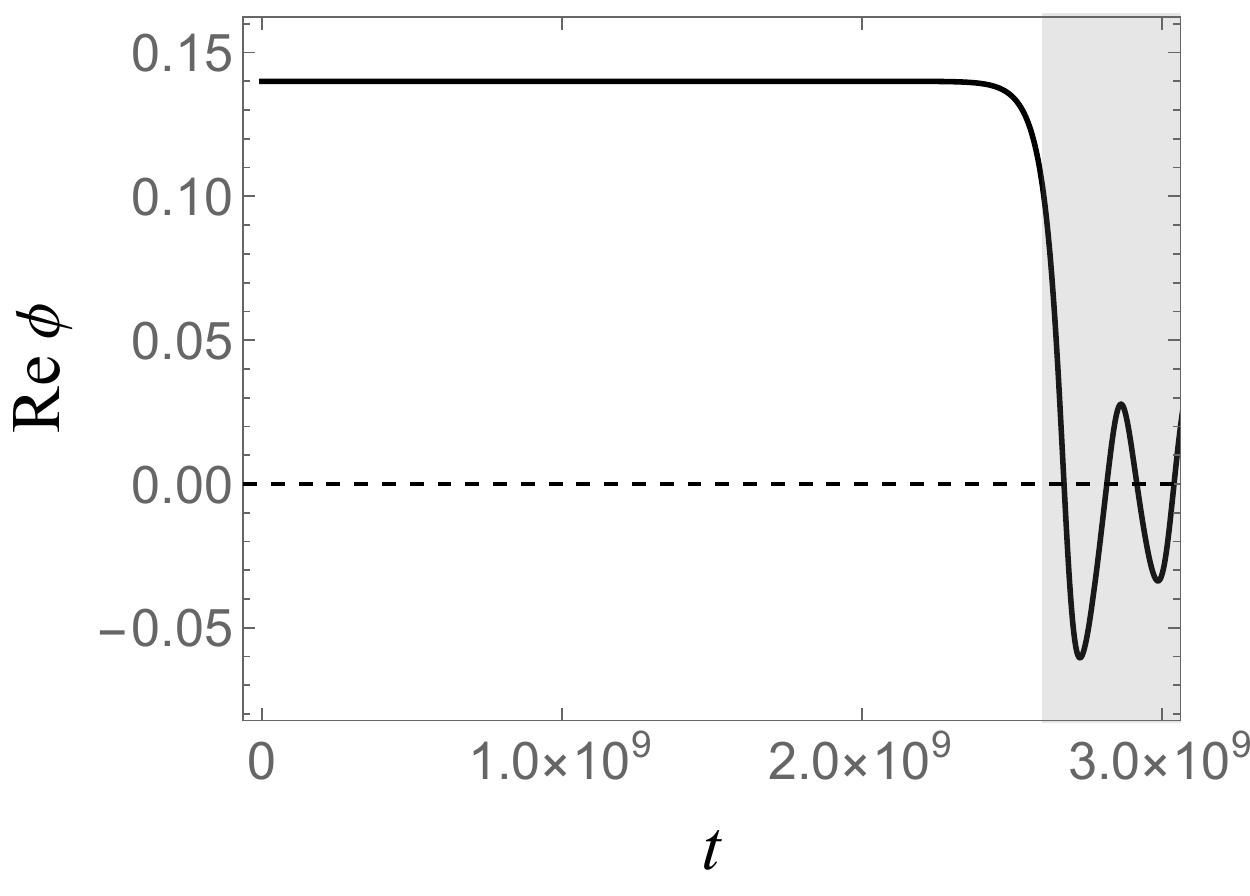}
		\caption{$\Re \phi$}
		\label{fig:cb}
	\end{subfigure}\hfill
	\begin{subfigure}[b]{0.33\textwidth}
		\centering
		\includegraphics[width=\linewidth]{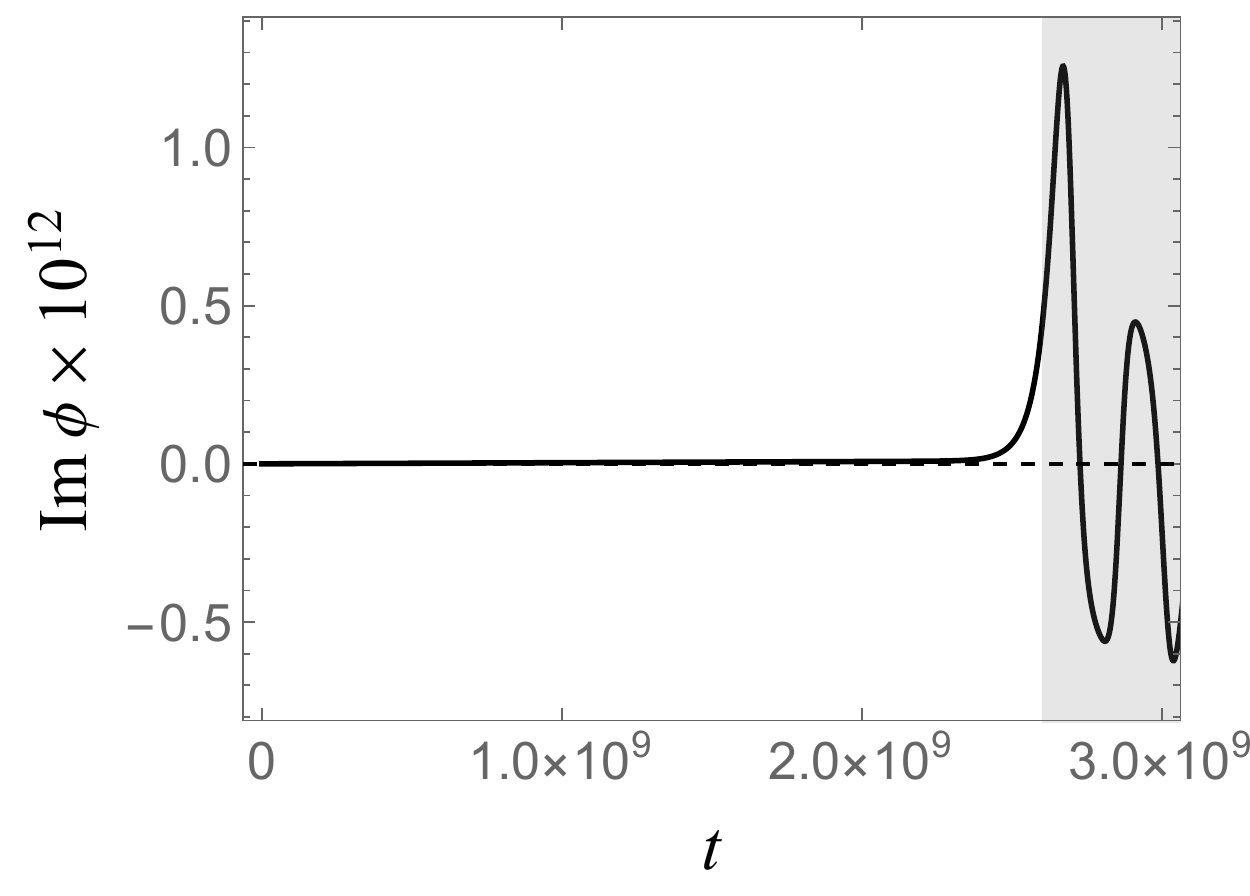}
		\caption{$\Im \phi$}
		\label{fig:cc}
	\end{subfigure}
	\begin{subfigure}[b]{0.33\textwidth}
		\centering
		\includegraphics[width=\linewidth]{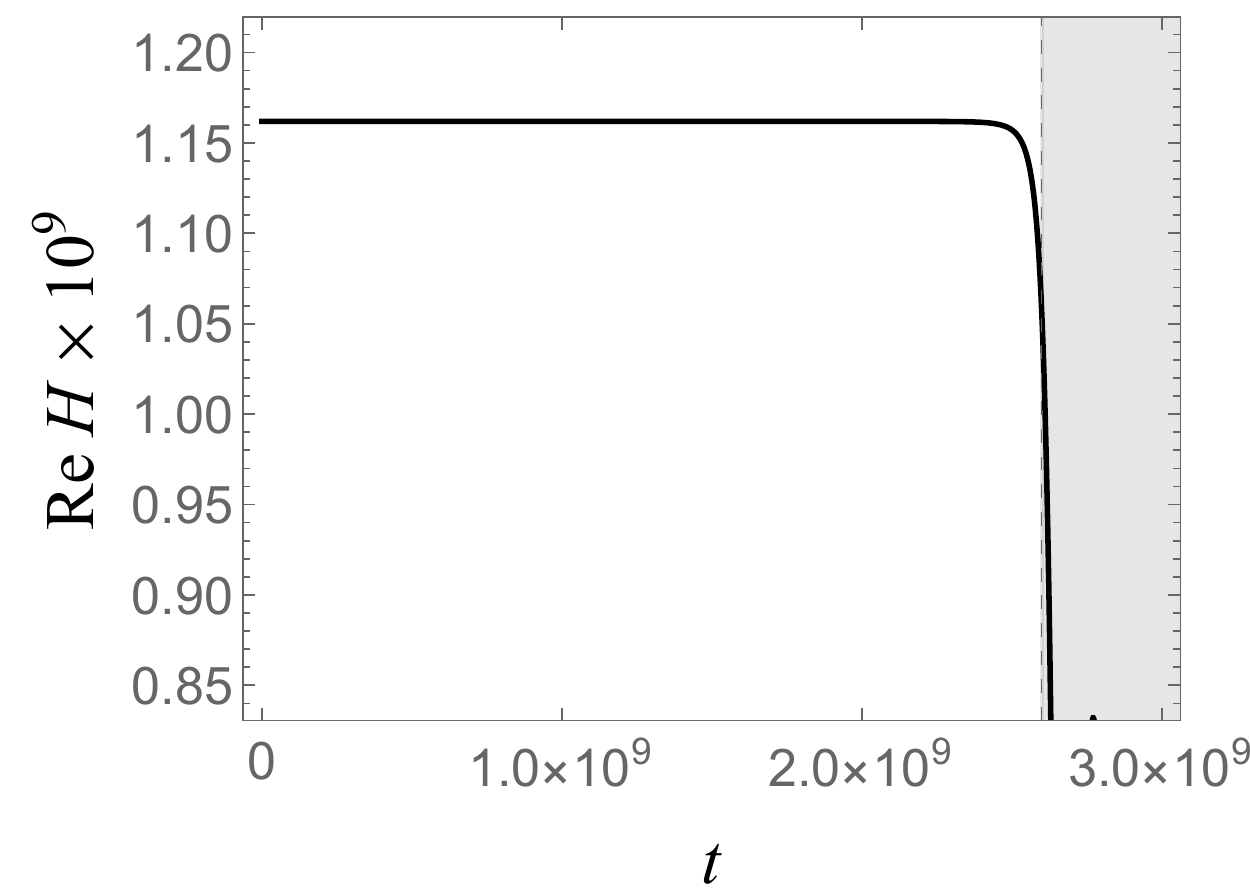}
		\caption{$\Re H$}
		\label{fig:1d}
	\end{subfigure}\hfill
	\begin{subfigure}[b]{0.33\textwidth}
		\centering
		\includegraphics[width=\linewidth]{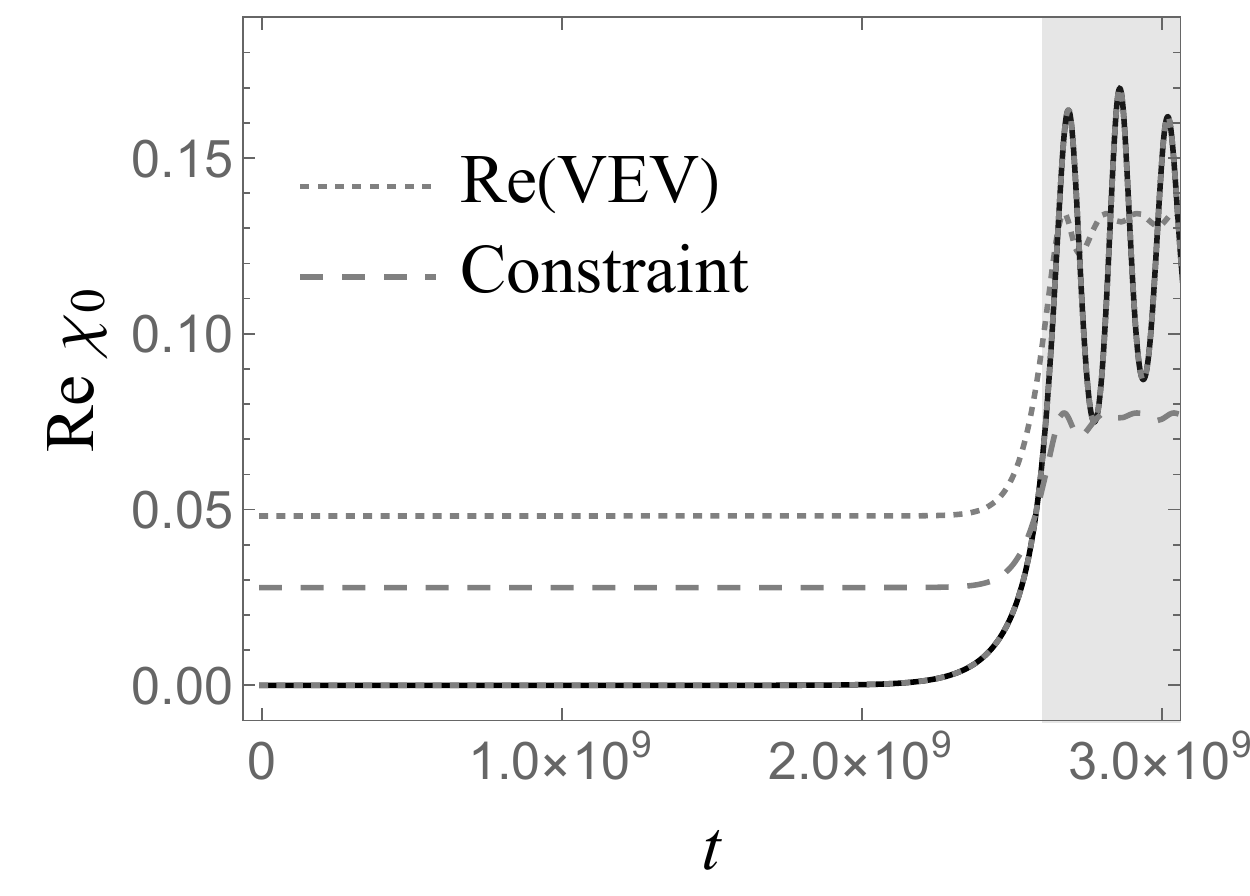}
		\caption{$\Re \chi_0$}
		\label{fig:1e}
	\end{subfigure}\hfill
	\begin{subfigure}[b]{0.33\textwidth}
		\centering
		\includegraphics[width=\linewidth]{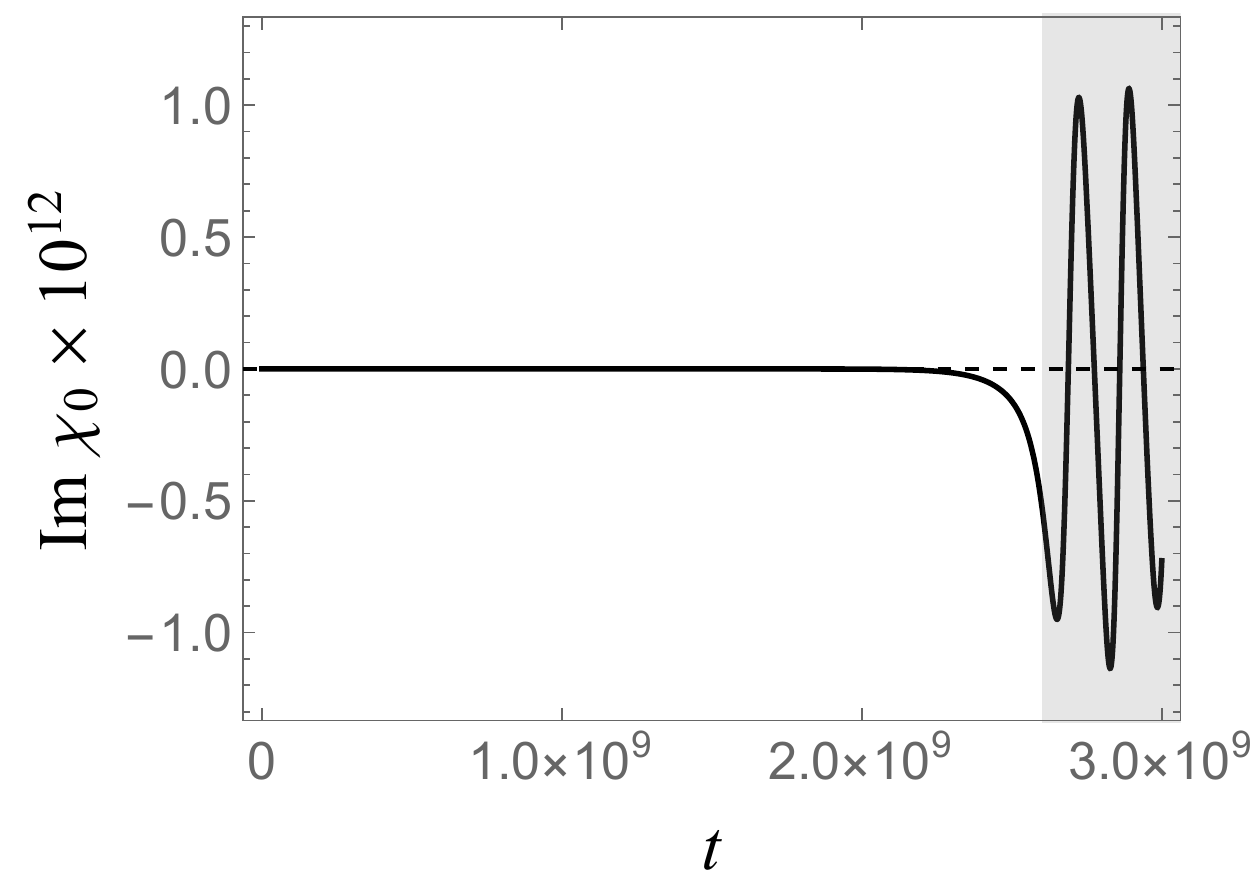}
		\caption{$\Im \chi_0$}
		\label{fig:1f}
	\end{subfigure}
	\caption{Plots of the slow-roll parameter $\epsilon$, the Hubble parameter and the fields for the parameter set \eqref{eq:n1a}. In the shaded region $\epsilon > 1$.}
\label{fig:clesse}
\end{figure}

We compute the system of equations \eqref{eq:i11}, \eqref{eq:i12} and \eqref{eq:i14} using the slow-roll initial conditions. The results can be seen in Figure~\ref{fig:clesse}. We find the waterfall field fluctuations have no effect during the first stage of the waterfall regime and they agree completely with the classical results, which can be obtained by computing only the classical system of equations of motion provided in Section~\ref{hyb}. Slow-roll inflation ends once the waterfall field reaches the steep part of the double-well potential, i.e. just before the symmetry breaking phase. The imaginary parts are very sub-dominant and therefore do not play any role in this case.

In Figure~\ref{fig:1e} we trace the constraint \eqref{eq:i2}. This indicates up to where the EFT is applicable. Indeed, as can be seen in this plot, we can extrapolate further than this point, but this is only because the waterfall field fluctuations are well controlled and the system obeys the classical dynamics. In truth, the theory, classical or otherwise, cannot be trusted further than where the constraint lies. This is where we enter the non-perturbative regime and other methods are needed to correctly compute the evolution of the system.

 The reason we extrapolate our results further than the validity of our EFT is to give the reader a feeling of where the effective potential becomes steep and where the waterfall field relaxes at its minimum. We have also shown how the vacuum expectation value of the waterfall field evolves during this process. Indeed, we see that initially, the evolution of the  waterfall field is well within the constraint and well below the vacuum expectation value, while the inflaton slow-rolls. Once the light waterfall modes start to grow the effective potential becomes too steep and slow-roll inflation ends, as it can be seen in Figure~\ref{fig:ca}. This is also the point where the EFT becomes invalid.

We find that as long as the waterfall backreaction remains under control, the theory is well described by the classical dynamics. We also display the evolution of the imaginary parts of the fields which are very small throughout and therefore have no effect, as expected.

\subsection{Standard hybrid inflation}

\begin{figure}
	\begin{subfigure}[b]{0.33\textwidth}
		\centering
		\includegraphics[width=\linewidth]{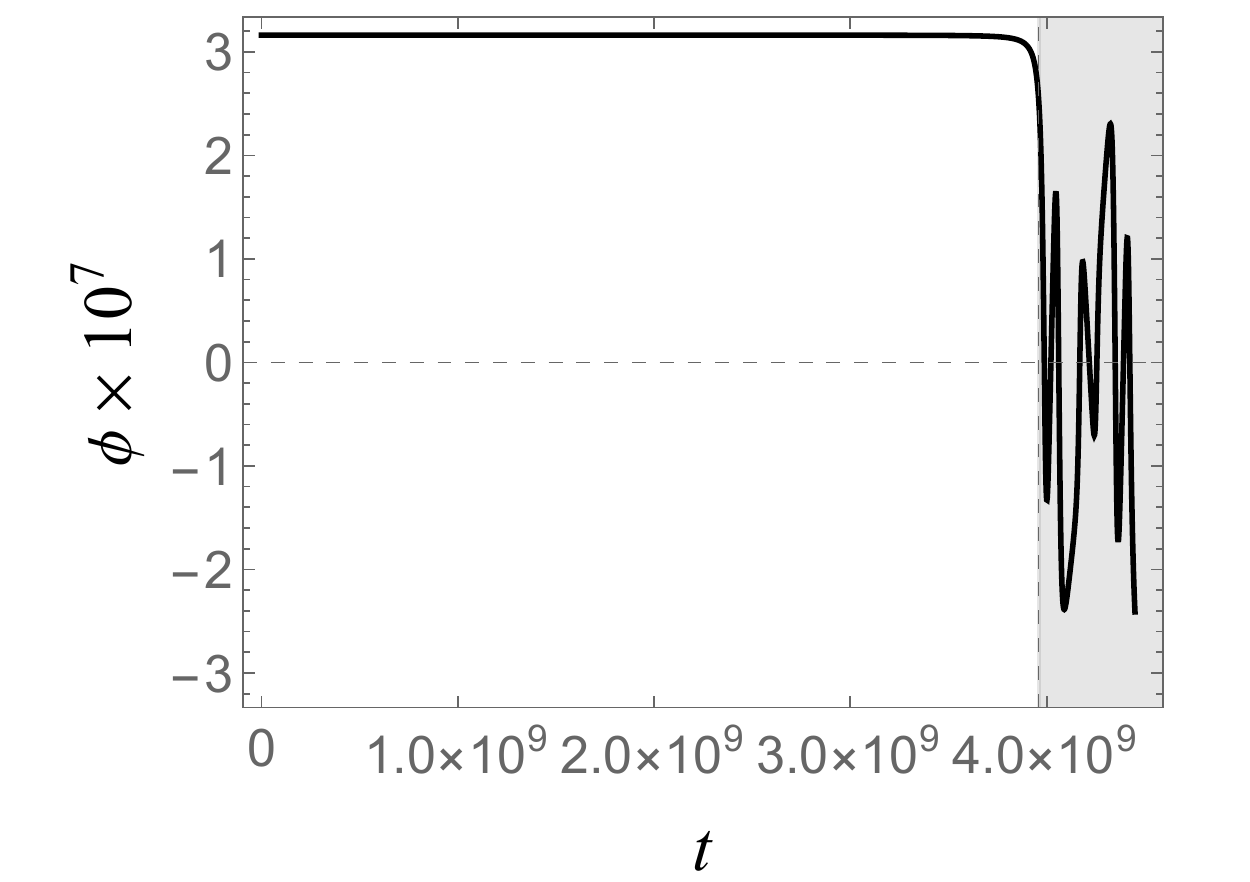}
		\caption{$ \phi$}
		\label{fig:l1a}
	\end{subfigure}\hfill
	\begin{subfigure}[b]{0.33\textwidth}
		\centering
		\includegraphics[width=\linewidth]{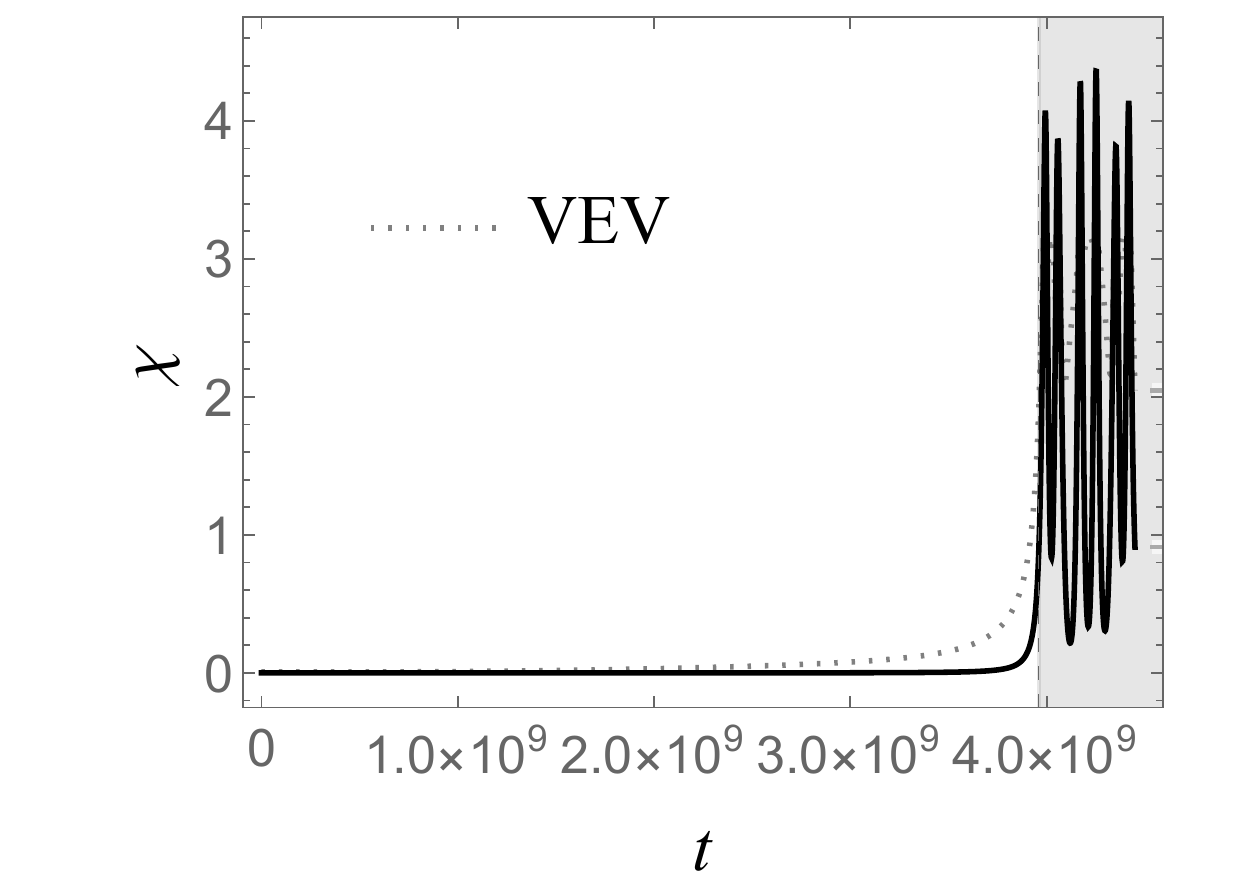}
		\caption{$ \chi$}
		\label{fig:l1b}
	\end{subfigure}\hfill
	\begin{subfigure}[b]{0.33\textwidth}
		\centering
		\includegraphics[width=\linewidth]{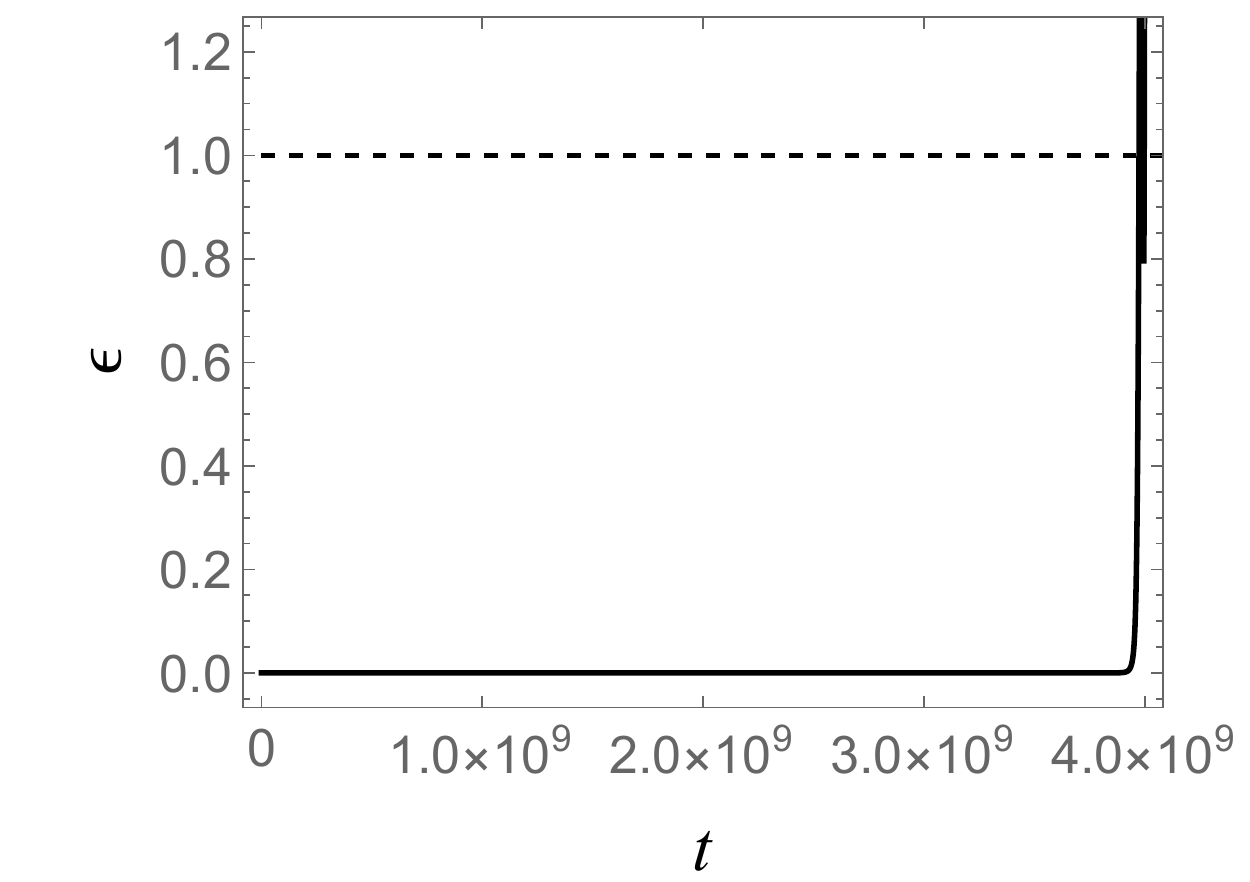}
		\caption{$ \epsilon$}
		\label{fig:l1c}
	\end{subfigure}
	\begin{subfigure}[b]{0.33\textwidth}
		\centering
		\includegraphics[width=\linewidth]{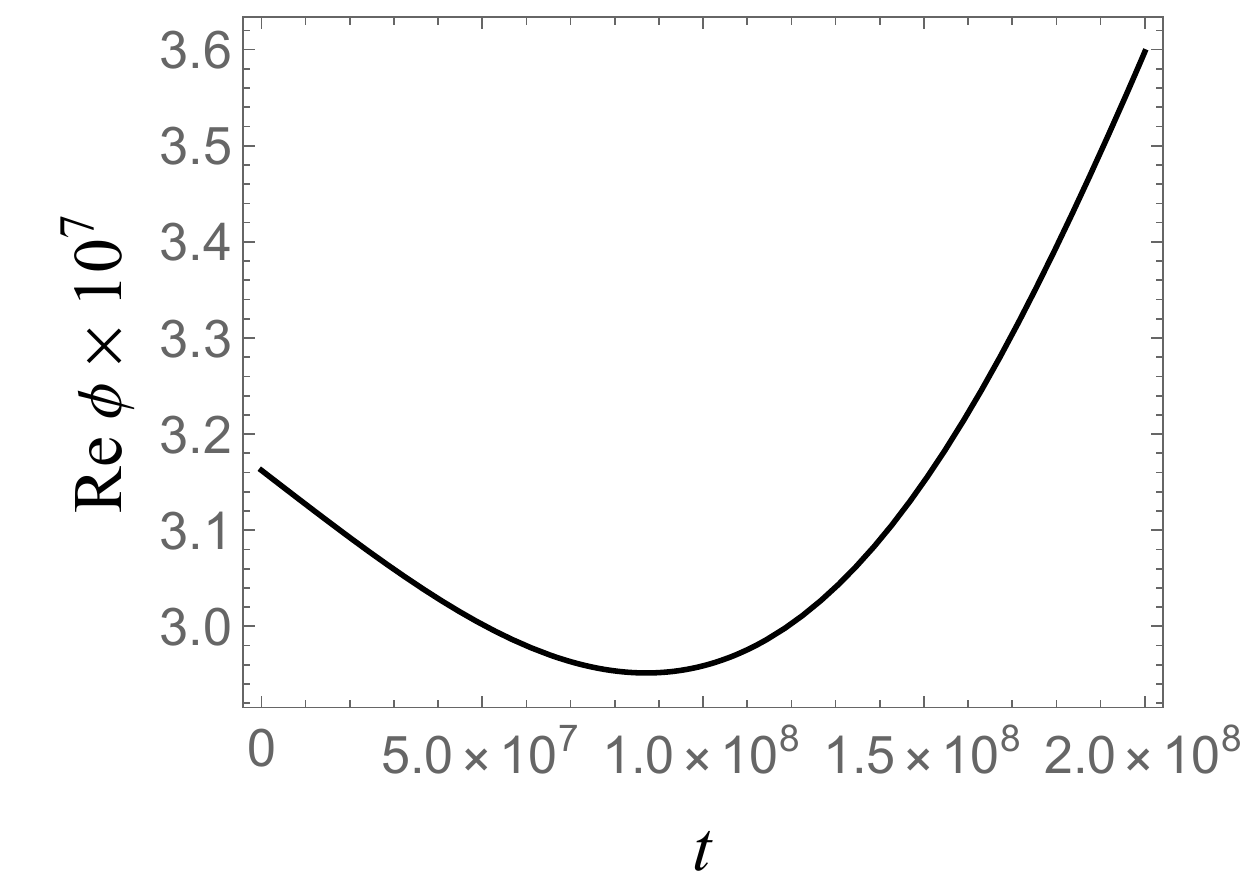}
		\caption{$\Re \phi$}
		\label{fig:l1d}
	\end{subfigure}\hfill
	\begin{subfigure}[b]{0.33\textwidth}
		\centering
		\includegraphics[width=\linewidth]{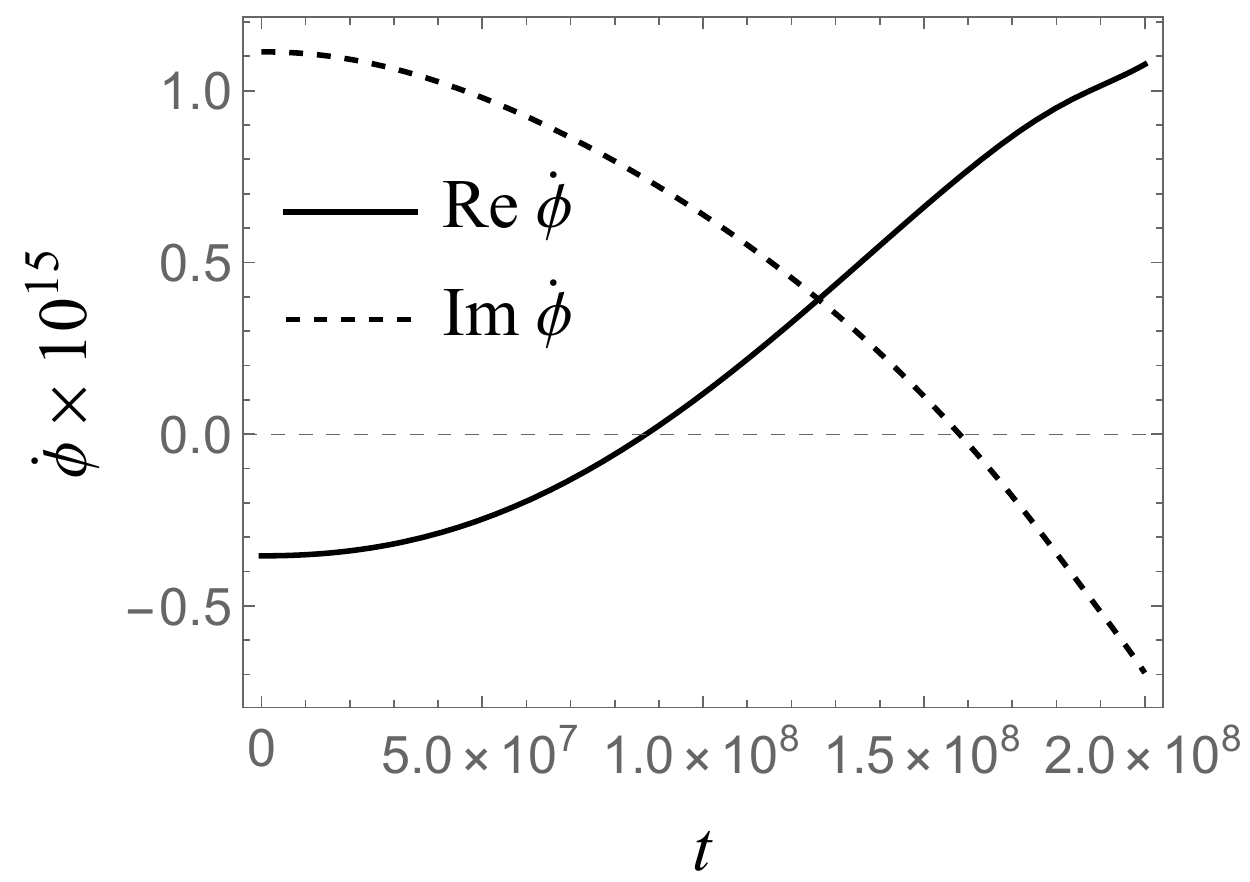}
		\caption{$\dot\phi$}
		\label{fig:l1e}
	\end{subfigure}\hfill
	\begin{subfigure}[b]{0.33\textwidth}
		\centering
		\includegraphics[width=\linewidth]{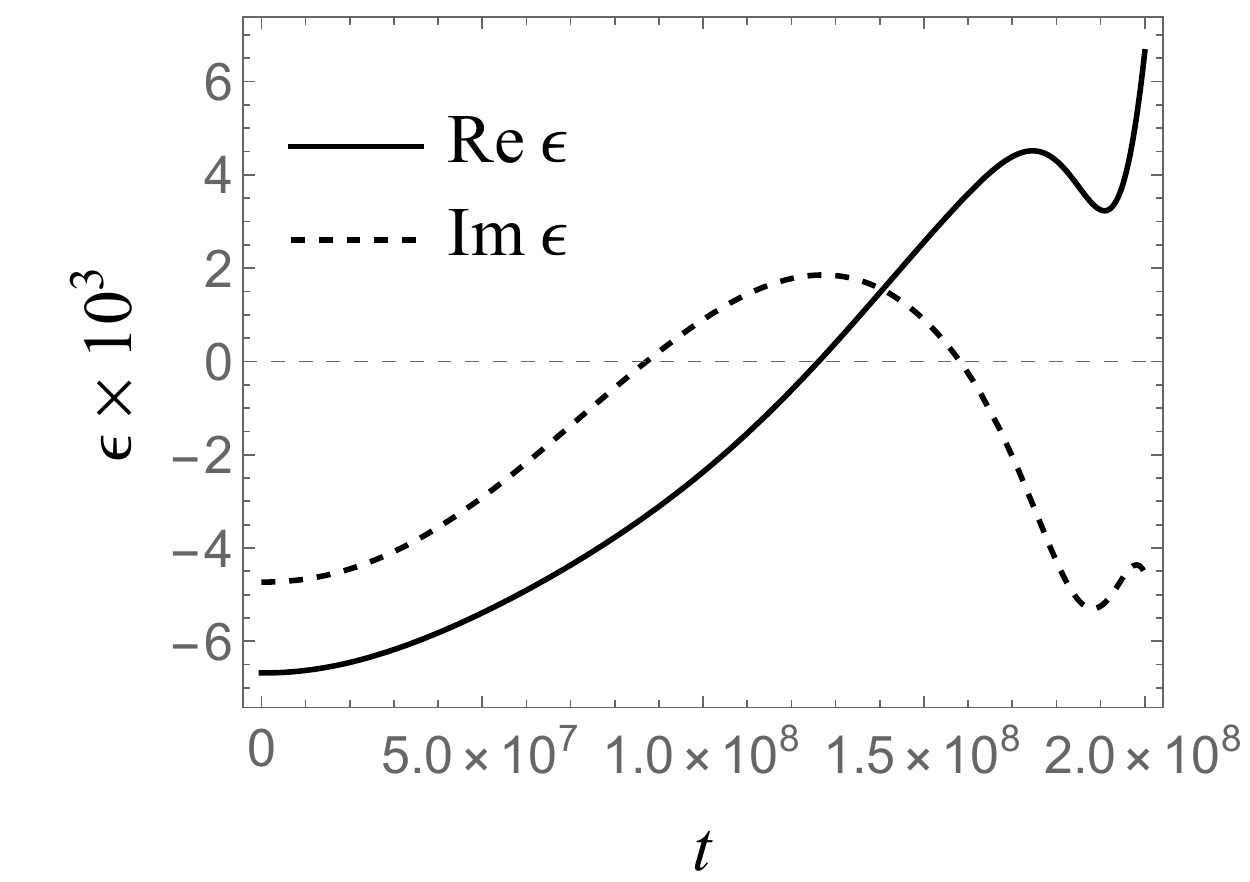}
		\caption{$\epsilon$}
		\label{fig:l1f}
	\end{subfigure}\hfill
	\begin{subfigure}[b]{0.33\textwidth}
		\centering
		\includegraphics[width=\linewidth]{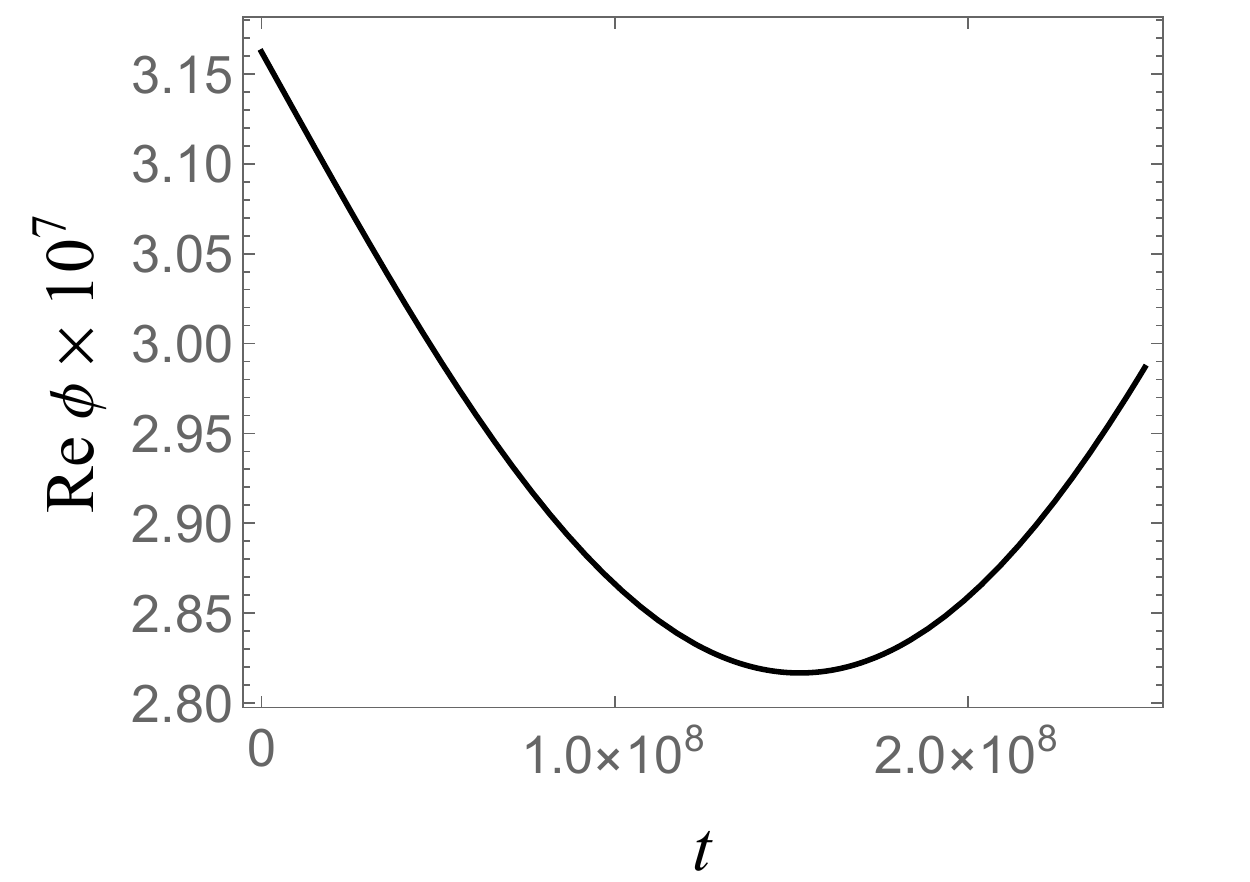}
		\caption{$\Re \phi$}
		\label{fig:l1g}
	\end{subfigure}\hfill
	\begin{subfigure}[b]{0.33\textwidth}
		\centering
		\includegraphics[width=\linewidth]{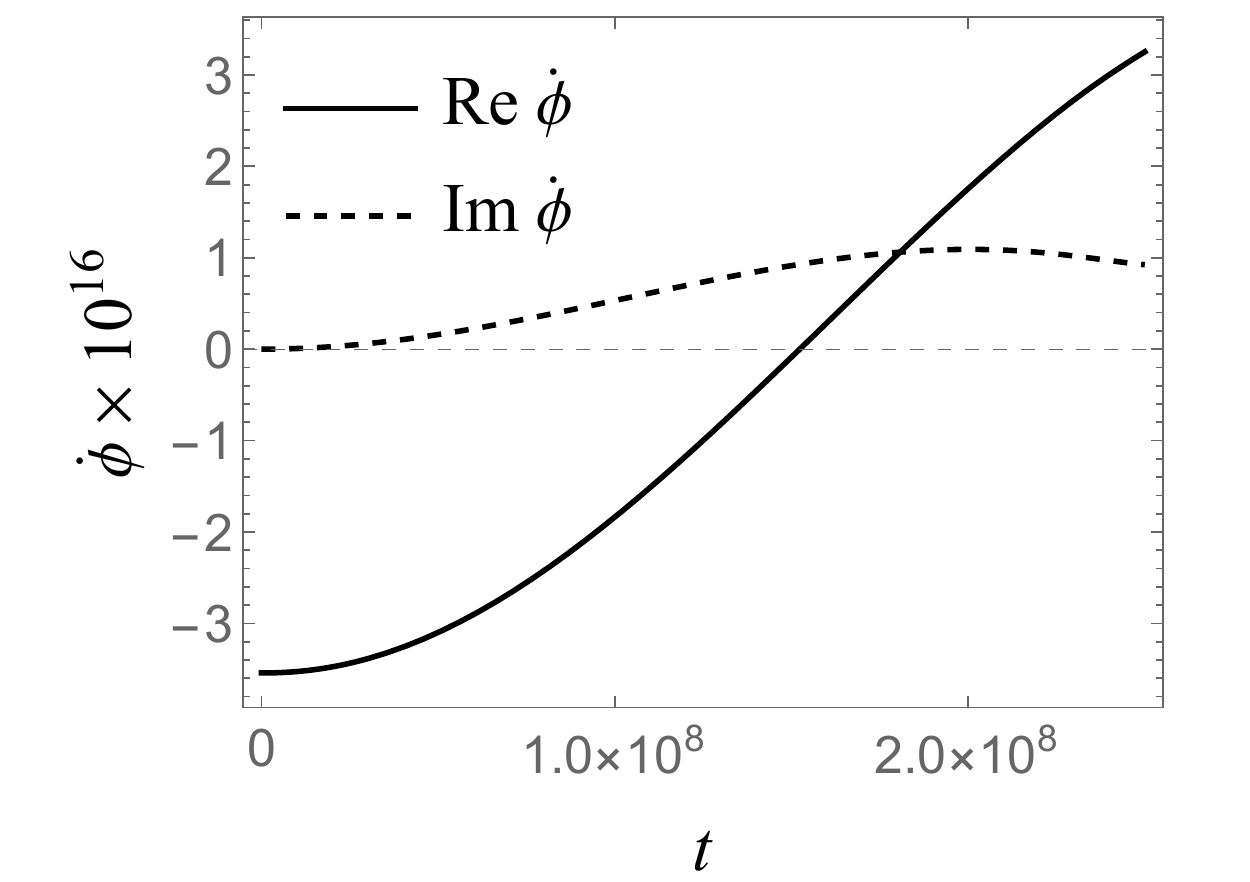}
		\caption{$\dot\phi$}
		\label{fig:l1h}
	\end{subfigure}\hfill
	\begin{subfigure}[b]{0.33\textwidth}
		\centering
		\includegraphics[width=\linewidth]{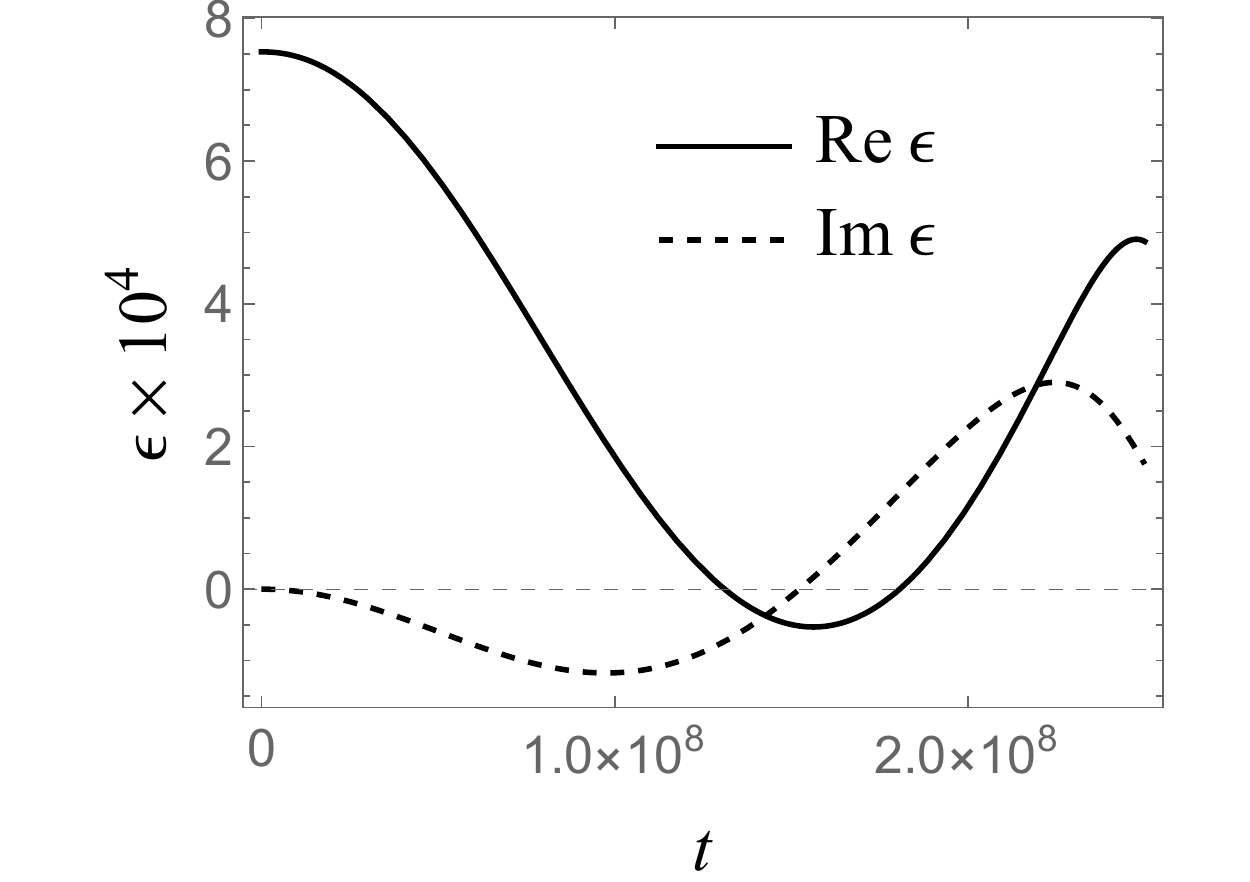}
		\caption{$\epsilon$}
		\label{fig:l1i}
	\end{subfigure}
	\caption{
		(Top row) Plots of the fields and the slow-roll parameter $\epsilon$ for the parameter set \eqref{eq:n1b}. In the shaded region $\epsilon > 1$.
		(Middle row) Plots of $\phi$, $\dot\phi$ and $\epsilon$ with quantum corrections with the same parameter set \eqref{eq:n1b}.
		(Bottom row) Plots of $\phi$, $\dot\phi$ and $\epsilon$ with quantum corrections, but setting $\Im\dot\phi(t=0)=0$ by hand.
	} \label{fig:linde1a}
\end{figure}

Next we consider the parameter set for standard hybrid inflation~\cite{Linde:1993cn} where the waterfall condition is obeyed. We use the first set of parameters (again in the unit $\mpl=1$) provided in \cite{Linde:1993cn} for which the waterfall transition is related to the electroweak symmetry breaking: 
\begin{equation}
\begin{split} 
& 
\lambda =  10^{-1}, \quad  g =  10^{-\frac{1}{2}}, \quad M= 10^{-7}, \quad m = 10^{-16}, 
\\
& 
\chi_0(t=0) = 10^{-14}, \quad \phi(t=0)= 3.15 \times 10^{-7}, \quad \Lambda \sim 2.2 \times 10^{-10}.
\label{eq:n1b} 
\end{split}
\end{equation}
The results can be seen in Figure~\ref{fig:linde1a}. In the first row we plot the classical evolution of the fields, i.e., using the classical equations of motion which can be found in Section~\ref{hyb}. We find slow-roll inflation ends at the onset of the symmetry breaking phase, in agreement with what has been demonstrated previously in the literature.

In the second row of Figure~\ref{fig:linde1a}, we plot the evolution of the $\phi$, $\dot\phi$ and $\epsilon$ in the presence of the quantum corrections, using \eqref{eq:i11}, \eqref{eq:i12} and \eqref{eq:i14}. The results indicate that we have lost completely the control of the EFT. This can be seen clearly in Figure~\ref{fig:l1e} as the velocity of the inflaton has a large imaginary part. This causes, consequently, $\epsilon$ to become negative as well to acquire a large imaginary part as shown in Figure~\ref{fig:l1f}. We should not take these results to be of actual physical relevance. These plots are only a visual demonstration of what it means to lose the validity of the EFT due to large backreaction from the waterfall field fluctuations. We would not expect such a situation to have happened physically.

The reason why the EFT is not valid at all is because with this set of initial conditions the mass of the inflaton is very small and therefore the potential very flat, as was discussed earlier. Meanwhile, the coupling constant $g$ is very large in comparison. Therefore, the constraint \eqref{eq:sl7} is violated so that from \eqref{eq:sl4} the inflaton velocity becomes complex from the beginning. This is because the quantum corrections, {\it both} real and imaginary, dominate from the onset of the waterfall phase, with the classical approximation being completely spoiled.

\begin{figure}
	\begin{subfigure}[b]{0.33\textwidth}
		\centering
		\includegraphics[width=\linewidth]{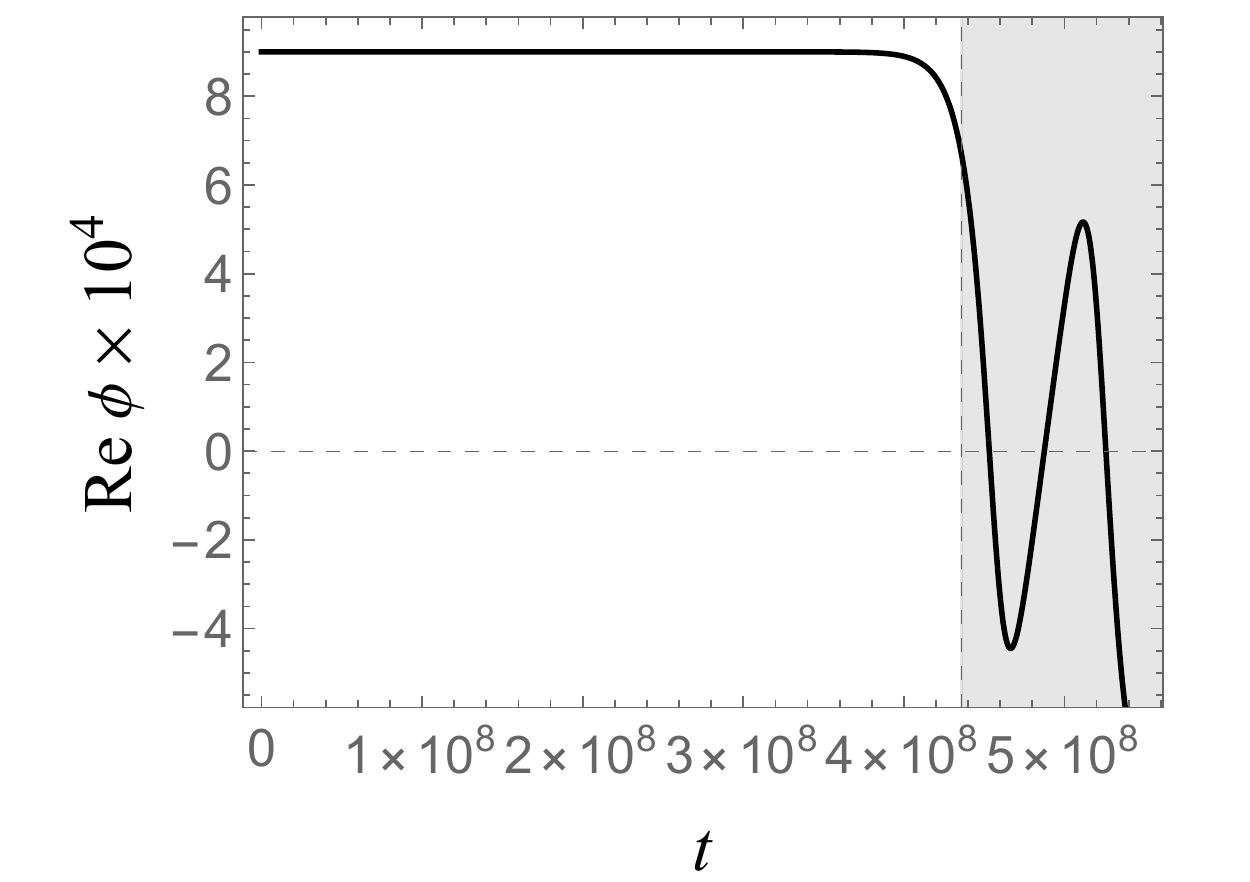}
		\caption{$\Re\phi$}
		\label{fig:1a}
	\end{subfigure}\hfill
	\begin{subfigure}[b]{0.33\textwidth}
		\centering
		\includegraphics[width=\linewidth]{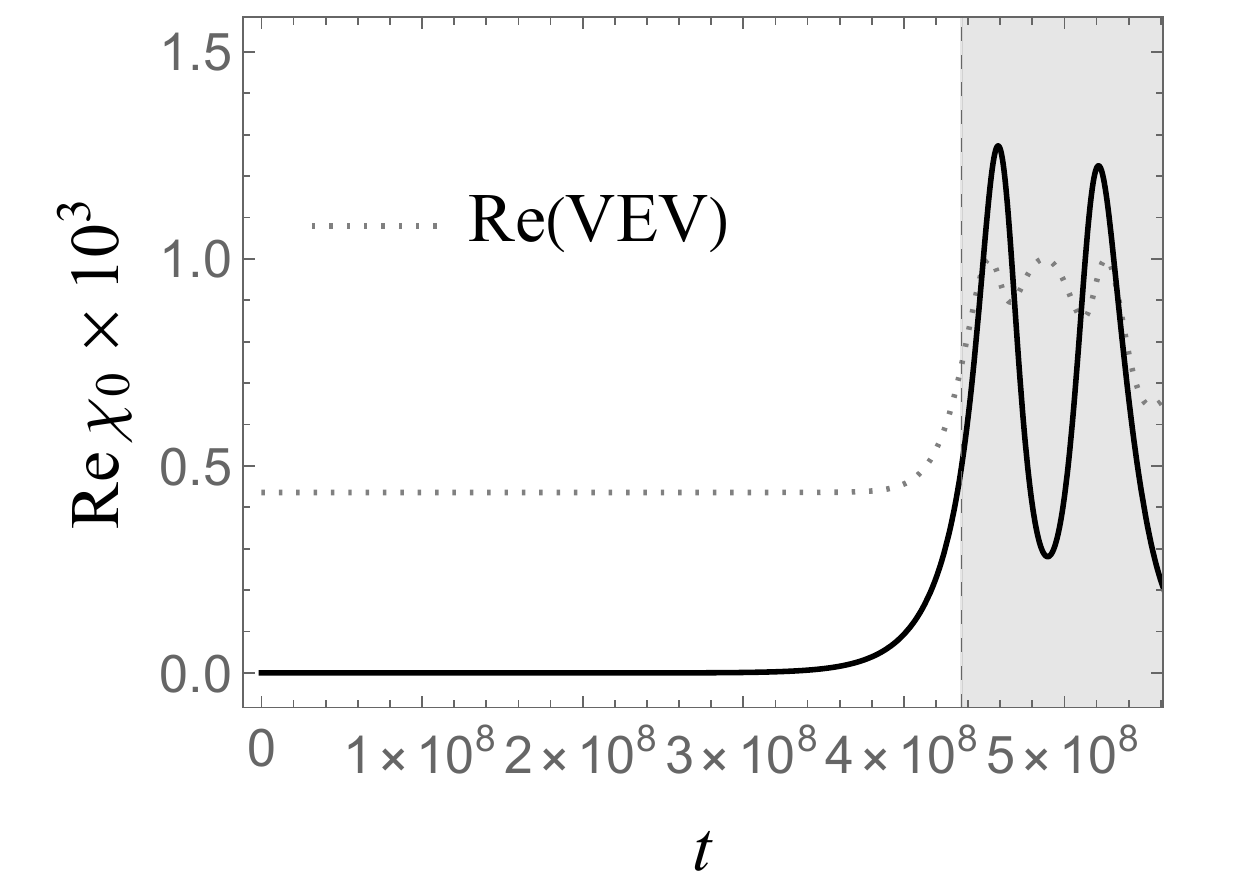}
		\caption{$\Re\chi_0$}
		\label{fig:1b}
	\end{subfigure}\hfill
	\begin{subfigure}[b]{0.33\textwidth}
		\centering
		\includegraphics[width=\linewidth]{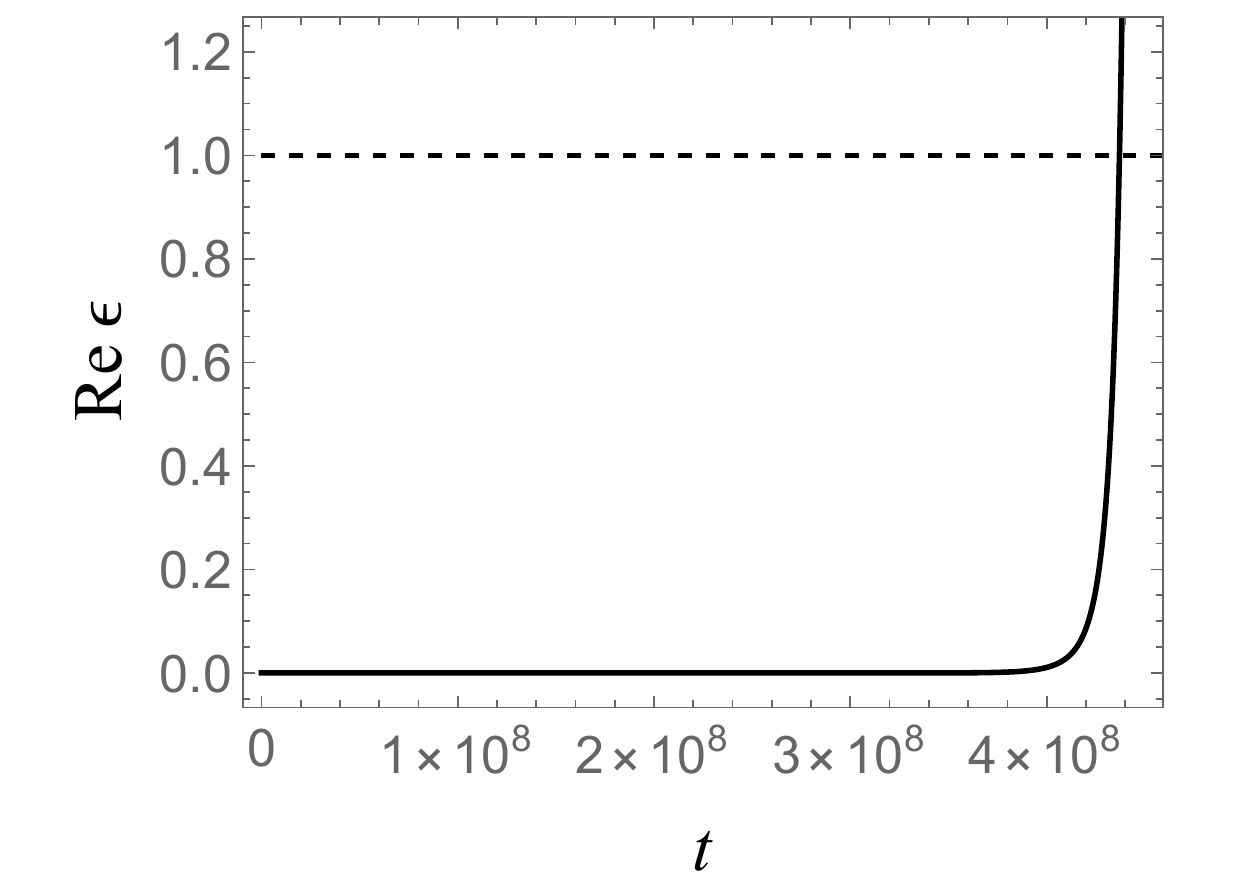}
		\caption{$\Re\epsilon$}
		\label{fig:1c}
	\end{subfigure}
	\caption{Plots of the fields and the slow-roll parameter $\epsilon$ for the parameter set \eqref{eq:n1d}. In the shaded region $\epsilon > 1$.} \label{fig:lindefix2}
\end{figure}

To make more visible the effect of the real part of the quantum corrections onto the system, we can naively set the imaginary part of the inflaton velocity to zero by hand. Indeed, the imaginary part of the inflaton velocity begins from zero but the real part soon evolves to reach the same order of magnitude as that of the imaginary part, as seen in Figure~\ref{fig:l1h}. Consequently, the sign flip of the real part of the inflaton velocity causes the inflaton to grow, as seen in Figure~\ref{fig:l1g}. From this we find that the EFT quickly becomes invalid again.  Therefore, we conclude that slow-roll inflation is, in the context of the EFT, not guaranteed to continue across the waterfall phase due to the large quantum backreaction.
We have also checked the second set of parameters provided in \cite{Linde:1993cn} regarding GUT phase transition, and found more or less the same behaviour. The parameters are
\begin{equation}
\begin{split} 
& 
\lambda =  1, \quad  g =  1, \quad M= 10^{-3}, \quad m =5\times 10^{-8}, 
\\
& 
\chi_0(t=0) = 10^{-14}, \quad \phi(t=0)= 9 \times 10^{-4}, \quad \Lambda \sim 4.3 \times 10^{-4}.
\label{eq:n1c} 
\end{split}
\end{equation}
This situation can be mended by imposing that the initial velocity of the inflaton must obey the constraint \eqref{eq:sl7}. To satisfy the constraint we need to increase the value of the mass of the inflaton and decrease the value of the coupling constant $g$. We choose the following parameters:
\begin{equation}
\begin{split} 
&
\lambda =  10^{-8}, \quad  g =  10^{-4}, \quad M= 10^{-7}, \quad m =9.9 \times 10^{-12}, 
\\
& 
\chi_0(t=0) = 10^{-14}, \quad  \phi(t=0)= 9 \times 10^{-4}, \quad \Lambda \sim 4.35 \times 10^{-8}
\label{eq:n1d} 
\end{split}
\end{equation}
Indeed, as we can see in Figure~\ref{fig:lindefix2}, this gives the behaviour expected to find during the waterfall regime. The waterfall field fluctuations are under control and the classical dynamics dominate. This is a good example of the importance of the constraint \eqref{eq:sl7} which warns that care must be taken when using the classical approximation. Finally, we mention that compared to our result for the evolution of the waterfall field, the naive expression for the growth of long-wavelength modes \eqref{eq:pp3b} gives an enhancement by a factor of $10^3$.

\section{Conclusions}
\label{sec:conc}

We have constructed the EFT of the waterfall transition and have derived a constraint \eqref{eq:sl7}, which must be satisfied in order to avoid strong backreaction of the waterfall field fluctuations onto the system during the early stage of the waterfall transition. This ensures the validity of the EFT and resulting classical approximation.
We find, what was previously thought as a viable parameter set in standard hybrid inflation, could violate this constraint, forcing slow-roll inflation to end abruptly at the onset of waterfall and making the classical approximation invalid.

This indicates the subtlety of the symmetry breaking processes in time-dependent backgrounds. The classical approximation has been used to study tachyonic preheating and generation of primordial black holes in various models of hybrid inflation. We find the waterfall field fluctuations may become important way before the symmetry breaking phase, i.e., at the onset of waterfall and cannot be neglected without a proper investigation. This may have important implications for the study of any spontaneous symmetry breaking dynamics in cosmology as well as subsequent phenomenologies.

\medskip

\subsection*{Acknowledgements}

We are indebted to Niayesh Afshordi and Cliff Burgess for their continuous encouragement 
and insightful correspondences.
We also thank Ki-Young Choi, Subodh Patil and Misao Sasaki for helpful discussions, 
and Benjamin L'Huillier for initial collaboration.
We are supported in part by the National Research Foundation of Korea Grant 2019R1A2C2085023.
JG also acknowledges the Korea-Japan Basic Scientific Cooperation Program supported 
by the National Research Foundation of Korea and the Japan Society for the Promotion of Science 
(2020K2A9A2A08000097). 
JG is further supported in part by the Ewha Womans University Research Grant of 2020 
(1-2020-1630-001-1) and 2021 (1-2021-1227-001-1). 
JG is grateful to the Asia Pacific Center for Theoretical Physics for hospitality 
while this work was under progress.

\appendix
\renewcommand{\theequation}{\Alph{section}.\arabic{equation}}

\section{Functional integration}
\label{app:Ap2} 
\setcounter{equation}{0}

The generating functional $W$ for connected Feynman diagrams is related to the partition function $Z$ by
\begin{equation}
Z[J]= \exp^{iW[J]} \quad \leftrightarrow \quad W[J] = -i \log Z[J] \, ,
\label{eq:017}
\end{equation}
where $J$ denotes an external  source. Differentiating once with respect to $J$, let us define the classical field $\chi_0$ by
\begin{equation}
\frac{\delta}{\delta J(x)} W[J] 
= i \frac{\delta}{\delta J(x)} \log Z 
= \frac{\int\mathcal{D}\chi \; \chi(x) \exp[i \int (\mathcal{L} + J \chi)]}
{\int\mathcal{D}\chi \; \exp[i \int (\mathcal{L} + J \chi)]} 
= \expval{\Omega}{\chi}_J 
\equiv \chi_0(x) \, .
\label{eq:018}
\end{equation}
We can define the effective action as the functional Legendre transform of $W[J]$: 
\begin{equation}
\Gamma[\chi_0] \equiv -W[J]-\int \dd^4{y} J(y)\chi_0(y) \, ,
\label{eq:019}
\end{equation}
where we can show that
\begin{equation}
\frac{\delta}{\delta \chi_0(x)} \Gamma[\chi_0] = - J(x) \, .
\label{eq:020}
\end{equation}
We can then express the effective potential $V_\text{eff}$ as
\begin{equation}
\Gamma(\chi_0) = - \text{Vol} \; V_\text{eff}(\chi_0) \, ,
\label{eq:022}
\end{equation}
where Vol denotes the volume of space-time. Then, through this relation, the condition that $\Gamma(\chi_0)$ has an extremum reduces to the simple equation
\begin{equation}
\frac{\dd}{\dd \chi_0}  V_\text{eff}(\chi_0) =0  \qquad (\chi_0 \neq 0) \, .
\label{eq:023}
\end{equation}
We can use this to determine the vacuum expectation value of the fields, while differentiating again can be used as a definition of the mass of the fields:
\begin{equation}
\frac{\pd^2}{\pd \chi_0^2}  V_\text{eff}(\chi_0) =m^2 \Bigg|_{\chi_0=0}
\quad \text{and} \quad
\frac{\pd^2}{\pd \chi_0^2}  V_\text{eff}(\chi_0) =m^2 \Bigg|_{\chi_0=\langle \chi \rangle}
\, ,
\label{eq:024}
\end{equation}
where we display the expressions for the cases with no symmetry breaking and with symmetry breaking, respectively.

The effective potential is a function of the classical fields therefore, it makes straightforward the study of quantum corrections to our system of equations. The benefit of working with the effective potential is that it can be used to describe all the interactions in the theory -- this includes the tree-level contributions plus quantum corrections.

\section{One-loop corrections to the waterfall potential}
\label{app:int-out} 
\setcounter{equation}{0}

The one-loop corrections to the waterfall potential are obtained by using the background field method, i.e. expanding the field around some classical value around the top of the  double-well potential $\langle \chi \rangle=0$ as
\begin{equation}
\chi = \chi_0 + \delta\chi \, ,
\label{eq:w5}
\end{equation}
and keeping only the terms quadratic in $\delta\chi$ -- the terms linear in $\delta\chi$ drop out because of the the background equations of motion, while the terms higher-order in $\delta\chi$ contribute to two-loops and higher so we do not consider them here. The Lagrangian for the waterfall field becomes

\begin{align}
\mathcal{L} 
& =  
-\frac{1}{2} (\pd_\mu \chi_0)^2 - \frac{1}{2} M_\text{eff}^2 \chi_0^2 - \frac{\lambda}{4} \chi_0^4 
+ \frac{1}{2} \qty[-(\pd_\mu \delta\chi)^2 - M_\text{eff}^2(\phi) \delta\chi^2 - 3\lambda \chi_0^2 \delta\chi^2]
\nonumber\\  
& =
\frac{1}{2} \delta\chi \qty[  \square -  M_\text{eff}^2(\phi) -3 \lambda \chi_0^2 ] \delta\chi
\nonumber\\
& \equiv
\frac{1}{2} \delta\chi D \delta\chi
\, ,
\label{eq:w7}
\end{align}
where the second expression is obtained after integrations by parts and $\square \equiv \pd_\mu \pd^\mu$. Thus, the Lagrangian has the form of the Klein-Gordon operator where we may define $M_\text{eff}^2(\phi) + 3 \lambda \chi_0^2 \equiv \mathfrak{m}^2$.

Using that, the generating functional is given by 
\begin{equation}
Z[J] = e^{i W[J]} =\int \mathcal{D} \chi e^{i S[\chi]} \, ,
\label{eq:w3}
\end{equation}
then, the generating functional for the waterfall field, expanded around the classical value $\chi = \chi_0 + \delta\chi$, gives an expression of the form:
\begin{equation}
Z[J] = e^{i W[J]} = \int \mathcal{D} \chi \exp 
\left\{i S[\chi_0] +  \frac{i}{2}\int \dd^4{x} \, \delta\chi D \delta\chi  \right\} \, .
\label{eq:ww2d}
\end{equation}
Therefore, the perturbative expansion considered here involves the propagator which is the inverse of $D$, defined by
\begin{equation}
\langle \delta\chi(k) \delta\chi(-k) \rangle = \frac{- i}{k^2+\mathfrak{m}^2} \, ,
\label{eq:ww2g}
\end{equation}
where we had taken a minus sign out of the expression of $D$ and expressed the differential operator in terms of momentum. Going to Euclidean space by defining $t= - i \tau$, we have
\begin{equation}
Z[J] = e^{W} = \int \mathcal{D} \chi \exp
\left\{ S[\chi_0] - \frac{1}{2} \int \dd^4{x} \, \delta\chi (-D_E) \delta\chi \right\}
\, ,
\label{eq:ww2e}
\end{equation}
where $D = D_E$. One can show that
\begin{equation}\begin{split} 
- \frac{1}{2} \int \dd^4{x} \,  \delta\chi D_E \delta\chi  = \det[ D_E]^{-\frac{1}{2}} \, ,
\label{eq:ww2b}
\end{split}\end{equation}
where we have used the standard formula for evaluating  Gaussian integrals. Therefore, the one-loop correction to the Euclidean action is given by
\begin{equation}
W 
=  
- \frac{1}{2} \Tr \log[-\square_E + \mathfrak{m}^2] 
= 
- \frac{1}{2} \sum_k \log \qty(k^2_E + \mathfrak{m}^2) 
= 
- \frac{1}{2} \text{Vol} \int \frac{\dd^4{k}_E}{(2\pi)^4} \log \qty( k^2_E + \mathfrak{m}^2 )
\, ,
\label{eq:ww2f}
\end{equation}
where we have evaluated the trace of the differential operator $\square$ as the sum of its eigenvalues and have expressed it as an integral over momenta multiplied by the volume of space-time. Here, the integral over momenta can be written as
\begin{equation}
- \frac{1}{2} \int \frac{\dd^4{k}_E}{(2\pi)^4} \log \qty( k^2_E + \mathfrak{m}^2 ) 
=  \frac{1}{2} \pd_\alpha \int \frac{\dd^4{k}_E}{(2\pi)^4} 
\frac{1}{(k_E^2+\mathfrak{m}^2)^\alpha} \bigg|_{\alpha= 0}
\, ,
\label{eq:w15}
\end{equation}
where we have used $\pd_\alpha (k_E^2+\mathfrak{m}^2)^{-\alpha} = -  (k_E^2+\mathfrak{m}^2)^{-\alpha} \ln(k_E^2+\mathfrak{m}^2)$ and set $\lim_{\alpha \rightarrow 0} (k_E^2+\mathfrak{m}^2)^{-\alpha} = 1$.

We proceed with dimensional regularization by employing the well-known formula \cite{Ramond:1981pw}
\begin{equation}
\int \frac{\dd^{2\omega}{l}}{(2\pi)^{2\omega}} \frac{1}{(l^2+M^2+2l p)^A} 
= 
\frac{\Gamma(A-\omega)}{(4\pi)^\omega \Gamma(A)} \frac{1}{(M^2-p^2)^{A-\omega}}
\, ,
\label{eq:w16}
\end{equation}
and set $\omega l = d$ for dimension, $M=\mathfrak{m}$ and $A=\alpha$. The integral then becomes
\begin{equation}
\frac{1}{2} \pd_\alpha \qty[\frac{\Gamma(\alpha-{d/2})}{(4\pi)^{d/2}\Gamma(\alpha)} 
\frac{1}{(\mathfrak{m}^2)^{\alpha-{d/2}}}] \Bigg|_{\alpha=0} 
= 
\frac{1}{2}  \frac{\Gamma(-{d/2})}{(4\pi)^{d/2}} (\mathfrak{m}^2)^{d/2} 
\, ,
\label{eq:w17}
\end{equation}
where in the last step we have differentiated and took the limit $\alpha \rightarrow 0$. Finally, using \eqref{eq:022}, the effective potential is expressed as 
\begin{equation}
V_\text{eff}(\chi) 
= 
- \frac{1}{\text{Vol}} \Gamma[\chi_0] 
= 
\frac{1}{2} M^2_\text{eff} \chi_0^2 +  \frac{\lambda}{4} \chi_0^4 
- \frac{1}{2}  \frac{\Gamma(-{d/2})}{(4\pi)^{d/2}} (\mathfrak{m}^2)^{d/2} + V_{ct}(\chi) \, ,
\label{eq:w19aa} 
\end{equation}
where $V_{ct}$ contains the counterterms we use to renormalise the theory:
\begin{equation}
V_{ct} =- \frac{1}{2} \delta_{M_\text{eff}} \chi_0^2 + \frac{1}{4} \delta_\lambda \chi_0^4 \, .
\label{eq:w19c} 
\end{equation}
The third term in \eqref{eq:w19aa} is divergent. Thus, we define $\varepsilon = 4-d$ and the following relations follow:
\begin{align}
x^\varepsilon 
& = 
e^{\varepsilon \log x} = 1+ \varepsilon \log x + \mathcal{O}(\varepsilon^2) \, ,
\label{eq:w22a} 
\\
\Gamma(-n+\varepsilon) 
& = 
\frac{(-1)^n}{n!} \qty[\frac{1}{\varepsilon}+\psi_1(n+1)+\mathcal{O}(\varepsilon)]
\, , 
\label{eq:w23a} 
\end{align}
where $\psi_1(n+1)= 1 + 1/2 + \cdots + 1/n - \gamma$ is the digamma function with $\gamma\approx0.577216$ being the Euler-Mascheroni constant. Then we Taylor expand and keep only terms up to $\mathcal{O}(\varepsilon)$. Thus, the divergent term can be written as
\begin{align}
\frac{\Gamma(-{d/2})}{(4\pi)^{{d/2}}} (\mathfrak{m}^2)^{{d/2}} 
& = 
\Gamma\qty(-2+\frac{\varepsilon}{2}) \frac{(4\pi)^{{\varepsilon/2}}}{(4\pi)^2} 
\frac{(\mathfrak{m}^2)^2}{(\mathfrak{m}^2)^{{\varepsilon/2}}} 
\nonumber\\
& =
\qty(\frac{1}{\varepsilon}+\frac{3}{4}-\frac{\gamma}{2}) 
\frac{1+\frac{\varepsilon}{2} \log(4\pi)}{(4\pi)^2} \mathfrak{m}^4 \qty[1-\frac{\varepsilon}{2} \log(\mathfrak{m}^2)]
\nonumber\\
& = 
\frac{1}{2} \frac{\mathfrak{m}^4}{(4\pi)^2} \qty[\frac{2}{\varepsilon} - \gamma - \log(\mathfrak{m}^2) +\frac{3}{2}]
\, .
\label{eq:w21a} 
\end{align}
It is now easy to see that it contains a simple pole. The divergence can be absorbed using appropriate renormalisation conditions as we will see below. Finally, substituting for $\mathfrak{m}^2 = M^2_\text{eff}+ 3 \lambda \chi_0^2$ gives
\begin{equation}
V_\text{eff}(\chi) 
=  
-\frac{1}{2} \abs{M^2_\text{eff}} \chi_0^2 +  \frac{\lambda}{4} \chi_0^4 
-\frac{1}{4} \frac{\qty( -\abs{M^2_\text{eff}}+ 3 \lambda \chi_0^2)^2}{(4\pi)^2} 
\qty[\frac{2}{\varepsilon} - \gamma - \log(\frac{ -\abs{M^2_\text{eff}}+ 3 \lambda \chi_0^2}{\Lambda^2}) 
+\frac{3}{2}] + V_{ct}(\chi),
\label{eq:w19d} 
\end{equation}
where we have made it explicit that we want to focus in the regime where the effective mass squared is negative, $- \abs{M_\text{eff}^2}<0$. The mass scale $\Lambda$ ensures that the logarithm is dimensionless. For small values of $\chi_0$ the logarithm becomes imaginary. As far as renormalisation is concerned we only need to focus on the real part. The imaginary part is finite and proportional to $ i \pi$.

This expression contains ultraviolet divergences which we need to subtract using the counterterms in $V_{ct}$ to ensure the theory is renormalised. For this purpose, we employ dimensional regularisation by using that the dimension $d$ can be expressed as $\varepsilon = 4-d$ and demanding that the effective potential $V_\text{eff}$ obeys the initial conditions for the classical configuration of the field. That is, we minimise for
\begin{equation}
\left. \frac{\dd^2 \Re V_\text{eff}}{\dd \chi_0^2} \right|_{\chi_0=0}= - \abs{ M^2_\text{eff}}
\quad \text{and} \quad
\left. \frac{\dd^4 \Re V_\text{eff}}{\dd \chi_0^4} \right|_{\chi_0=0}= 6 \lambda
\, .
\label{eq:w19f} 
\end{equation}
From here onwards we omit that we are working with the real part. Now, using the effective potential \eqref{eq:w19aa} we solve for conditions \eqref{eq:w19f}, this gives
\begin{align}
\delta_{M_\text{eff}} 
& = 
3 \lambda \abs{ M^2_\text{eff}} \frac{\Gamma\qty(2-{d/2})}{(4\pi)^{d/2}} + \mathcal{O}(\varepsilon)
\, ,
\\
\delta_\lambda 
& = 
9 \lambda^2 \frac{\Gamma\qty(2-{d/2})}{(4\pi)^{d/2}} + \mathcal{O}(\varepsilon) 
\, ,
\label{eq:rc7a} 
\end{align}
where we have used  $\Gamma(n+1) = n \Gamma(n)$ for $n=-d/2$. Using $\varepsilon = 4-d$ and relations \eqref{eq:w22a} and \eqref{eq:w23a}, we find that the counterterms cancel the simple poles in \eqref{eq:w19d}. These have the form
\begin{equation}
- \frac{3 \lambda \abs{ M^2_\text{eff}} \chi_0^2}{16 \pi^2 \varepsilon} 
+\frac{9 \lambda^2  \chi_0^4}{32 \pi^2 \varepsilon} 
+\frac{ \abs{ M^4_\text{eff}} }{32 \pi^2 \varepsilon} 
\, .
\label{eq:rc11} 
\end{equation}
The first pole is removed with the $\delta_{M_\text{eff}}$ counterterm, while the second pole is removed with the $\delta_\lambda$ counterterm. The last term is not dynamical as it does not depend on $\chi_0$, thus it is just a shift of the effective potential and can always be absorbed in the normalisation of the path integral. Note that although it depends on the inflaton $\phi$ which is dynamical, the value of $\phi$ is taken to be almost constant during slow-roll inflation, i.e. in the cases examined here, so this is fine.

The finite terms can be combined with the finite terms coming from the counterterms to give a constant term, which we ignore as it is not dynamical. Finally, the effective potential becomes
\begin{equation}
V_\text{eff}(\chi) 
= 
- \frac{1}{2}\abs{ M^2_\text{eff}} \chi_0^2 +  \frac{\lambda}{4} \chi_0^4 
+\frac{1}{4} \frac{\qty(-\abs{M_\text{eff}^2} + 3 \lambda \chi_0^2)^2}{(4\pi)^2 } 
\qty[\log\qty(\frac{ -\abs{M_\text{eff}^2(\phi) }+ 3 \lambda \chi_0^2}{\Lambda^2})-\frac{3}{2}]
\, .
\label{eq:rc14} 
\end{equation}
This is the effective potential used in Section~\ref{sec:Veff}.

\medskip

\section{Small logs approximation}
\label{app:Taylor-log} 
\setcounter{equation}{0}

There will be times when the logarithms contained in these expressions complicate further calculations. To avoid this we can Taylor expand the logarithm in $V_\text{eff}$. This can be achieved if we demand that the initial value of the log argument is unity, which ensures small logs as part of our initial conditions. From this we find  good agreement with the original expression near the transition point but the computation will become inaccurate as the inflaton rolls further down in the potential.  In the cases considered here this won't play much of a role as the inflaton is almost constant during slow-roll inflation.  We find  good numerical agreement with the original and Taylor-expanded expressions as long as we stay within the perturbative regime of the theory.

Therefore, in order to simplify our arguments, we make the following approximation. We make a Taylor expansion for $x\ll1$ of the form
\begin{equation}
\log x = \sum_{n=1}^\infty \frac{(-1)^{n-1}(x-1)^n}{n} = (x-1) + \mathcal{O}(x^2)
\, .
\label{eq:i4} 
\end{equation}
The effective potential then becomes
\begin{equation}
V_\text{eff}(\chi_0) 
= 
- \frac{1}{2}   \abs{ M_\text{eff}}^2 \chi_0^2 +  \frac{\lambda}{4} \chi_0^4 
+\frac{1}{4} \frac{\qty( - \abs{ M_\text{eff}}^2 + 3 \lambda \chi_0^2)^2}{(4\pi)^2 } 
\qty[i\pi+\abs{\frac{ \abs{ M_\text{eff}}^2 - 3 \lambda \chi_0^2 }{\Lambda^2}}-\frac{5}{2}]
\, .
\label{eq:i5} 
\end{equation}
This way we can avoid dealing with the fact that the log blows up for some value of $\chi_0$ [see the constraint \eqref{eq:i2}] and focus on the regime of interest, which is the quantum instability due to the imaginary contribution in \eqref{eq:i5}, namely
\begin{equation}
\frac{i \pi}{4} \frac{\qty( - \abs{ M_\text{eff}} + 3 \lambda \chi_0^2)^2}{(4\pi)^2 } \, .
\label{eq:i6} 
\end{equation}
The correction from the quantum instability is, again, small as expected for the perturbation theory to be valid. Finally we can write our system of equations as follows:

\begin{itemize}

\item 
The inflaton equation takes the form
\begin{equation}
\ddot\phi+3H\dot\phi+ V_{\text{eff},\phi}=0 \, ,
\label{eq:qc1c}
\end{equation}
where
\begin{equation}
V_{\text{eff},\phi} 
= 
\qty(\mu_1^2 + \mu_2^2 \chi_0^2+\mu_3^2 \chi_0^4)\phi 
+ \qty(\nu_1 + \nu_2 \chi_0^2) \phi^3 + \nu_3 \phi^5
\, .
\label{eq:qc1}
\end{equation}
Here the parameters take complex values:
\begin{equation}
\begin{split} 
& 
\mu_1^2 = m^2 + \frac{5 g^2 M^2}{32 \pi^2} - \frac{3 g^2 M^4}{32 \pi^2 \Lambda^2} 
- \frac{ i g^2 M^2}{16 \pi}, 
\quad 
\mu_2^2 = g^2 + \frac{15 g^2 \lambda}{32 \pi^2} +\frac{9 g^2 M^2 \lambda}{16 \pi^2 \Lambda^2} 
+ \frac{3 i g^2 \lambda}{16 \pi},
\quad
\mu_3^2 = - \frac{27 g^2 \lambda^2}{32 \pi^2 \Lambda^2},
\\
& 
\nu_1 = -\frac{5 g^4}{32 \pi^2} + \frac{3 g^4 M^2}{16 \pi^2 \Lambda^2} + \frac{i g^4}{16 \pi}, 
\quad 
\nu_2 = -\frac{9 g^4 \lambda}{16 \pi^2 \Lambda^2}, 
\quad 
\nu_3 = -\frac{3 g^6}{32 \pi^2 \Lambda^2}.
\label{eq:qc1a}
\end{split}
\end{equation}

\item 
The waterfall field equation of motion becomes
\begin{equation}
\ddot\chi_0+3H\dot\chi_0 +V_{\text{eff},{\chi_0}} = 0 \, .
\label{eq:qc3a}
\end{equation}
Here,
\begin{equation}
V_{\text{eff},{\chi_0}} 
= 
\qty( \alpha_1 \abs{ M^2_\text{eff}}+ \alpha_2  \abs{ M^2_\text{eff}}^2) \chi_0 
+ \lambda \qty(\alpha_3 + \alpha_4 \abs{ M^2_\text{eff}}) \chi_0^3 + \alpha_5 \chi_0^5 \, ,
\label{eq:qc3}
\end{equation}
where
\begin{equation}
\alpha_1 =- 1 + \frac{15 \lambda}{32 \pi^2} - \frac{3 i \lambda}{16 \pi}, 
\quad 
\alpha_2= - \frac{9 \lambda }{32 \pi^2 \Lambda^2}, 
\quad 
\alpha_3 = 1 - \frac{45 \lambda}{32 \pi^2} + \frac{9 i \lambda}{16 \pi}, 
\quad 
\alpha_4 = \frac{27 \lambda}{16 \pi^2 \Lambda^2}, 
\quad 
\alpha_5 =  - \frac{81 \lambda^3}{32 \pi^2 \Lambda^2}.
\label{eq:qc4}
\end{equation}

\item
Finally, the Friedmann equation becomes
\begin{equation}
H^2 = \frac{1}{3\mpl^2} 
\qty[\frac{\dot\phi^2}{2}+ \frac{\dot\chi_0^2}{2} 
+ \frac{M^4}{4 \lambda}+ \frac{1}{2} m^2 \phi^2 +V_\text{eff}(\phi,\chi_0)]
\, ,
\label{eq:qc5}
\end{equation}
where
\begin{equation}
V_\text{eff}(\phi,\chi_0) 
= 
- \frac{1}{2}   \abs{ M^2_\text{eff}} \chi_0^2 +  \frac{\lambda}{4} \chi_0^4 
+ \frac{1}{4} \frac{\qty( - \abs{ M^2_\text{eff}} + 3 \lambda \chi_0^2)^2}{(4\pi)^2 } 
\qty[i\pi+ \frac{ \abs{ M^2_\text{eff}}^2 - 3 \lambda \chi_0^2 }{\Lambda^2}-\frac{5}{2}]
\, .
\label{eq:i7} 
\end{equation}

\end{itemize}

\bibliography{bibi} 
\bibliographystyle{utphys}

\end{document}